\documentclass[mnsc,nonblindrev]{informs3} 

\OneAndAHalfSpacedXI

\usepackage{bbm}

\usepackage{natbib}
 \bibpunct[, ]{(}{)}{,}{a}{}{,}%
 \def\bibfont{\small}%
\TheoremsNumberedThrough     
\ECRepeatTheorems

\EquationsNumberedThrough    

\MANUSCRIPTNO{}

\begin{document}


 \RUNAUTHOR{Holtz et al.}

\RUNTITLE{Reducing Bias from Interference in Online Marketplace Pricing Experiments}

\TITLE{Reducing Interference Bias in Online Marketplace Pricing Experiments}

\ARTICLEAUTHORS{%
\AUTHOR{David Holtz}
\AFF{MIT Sloan School of Management, Cambridge, MA 02142, \EMAIL{dholtz@mit.edu}} 
\AUTHOR{Ruben Lobel, Inessa Liskovich}
\AFF{Airbnb Inc., San Francisco, CA 94103, \EMAIL{ruben.lobel@airbnb.com} \EMAIL{inessa.liskovich@airbnb.com}}
\AUTHOR{Sinan Aral}
\AFF{MIT Sloan School of Management, Cambridge, MA 02142, \EMAIL{sinan@mit.edu}} 
} 

\ABSTRACT{%
Online marketplace designers frequently run A/B tests to measure the impact of proposed product changes. However, given that marketplaces are inherently connected, total average treatment effect estimates obtained through Bernoulli randomized experiments are often biased due to violations of the stable unit treatment value assumption. This can be particularly problematic for experiments that impact sellers' strategic choices, affect buyers' preferences over items in their consideration set, or change buyers' consideration sets altogether. In this work, we measure and reduce bias due to interference in online marketplace experiments by using observational data to create clusters of similar listings, and then using those clusters to conduct cluster-randomized field experiments. We provide a lower bound on the magnitude of bias due to interference by conducting a meta-experiment that randomizes over two experiment designs: one Bernoulli randomized, one cluster randomized. In both meta-experiment arms, treatment sellers are subject to a different platform fee policy than control sellers, resulting in different prices for buyers. By conducting a joint analysis of the two meta-experiment arms, we find a large and statistically significant difference between the total average treatment effect estimates obtained with the two designs, and estimate that 32.60\% of the Bernoulli-randomized treatment effect estimate is due to interference bias. We also find weak evidence that the magnitude and/or direction of interference bias depends on extent to which a marketplace is supply- or demand-constrained, and analyze a second meta-experiment to highlight the difficulty of detecting interference bias when treatment interventions require intention-to-treat analysis.
}%



\maketitle

%

\section{Introduction}

As of 2020, some of the world's most highly valued technology firms (e.g., Airbnb, Uber, Etsy) are online peer-to-peer marketplaces. These platforms create markets for many different types of goods, including accommodations, transportation, artisanal goods, and dog walking. Like almost all technology firms, online peer-to-peer marketplaces typically rely on experimentation, or A/B testing, to measure the impact of proposed changes to the platform and develop a deeper understanding of their customers. However, a randomized experiment's ability to provide an unbiased estimate of the total average treatment effect (TATE) relies on the stable unit treatment value assumption (SUTVA) \citep{rubin.1974}, sometimes referred to as the ``no interference" assumption \citep{cox.1958}. Online marketplaces are inherently connected; sellers are likely to make strategic decisions based on the actions of their competitors, and multiple sellers may sell different items that complement or substitute for one another. As a result, SUTVA is unlikely to hold in online marketplace settings. Previous work \citep{blake.2014, fradkin.2015, holtz.2018} has shown that naive experimentation in online marketplaces can lead to TATE estimates that are overstated by up to 100\%.

SUTVA violations are not unique to online marketplaces, and are a familiar problem for researchers conducting experiments in networked settings (e.g., social network experiments). In the network experimentation literature, researchers have proposed experiment designs \citep{eckles.2017, ugander.2013} and analysis techniques \citep{aronow.2012, eckles.2017} that aim to reduce bias due to statistical interference (henceforth referred to as interference bias), and \cite{saveski.2017} describe a procedure for ``randomizing over randomized experiments," or running meta-experiments, to detect interference bias on networks. \cite{holtz.2018} proposes the use of bias-reduction techniques from the networks literature to reduce bias in online marketplace experiments, and investigates the viability of this approach through a simulation study using scraped Airbnb data. However, this approach has, as of yet, not been used in the field to conduct randomized experiments in online marketplaces.

In this paper, we present the results from two meta-experiments conducted on Airbnb, an online marketplace for sharing homes and experiences. Both meta-experiments make use of clusters of Airbnb listings, which are created by first using observational search behavior to create a 16-dimension ``demand embedding" for each each Airbnb listing, and then segmenting the listing embedding space using a recursive partitioning tree. Each meta-experiment randomly assigns clusters of Airbnb listings to one of two randomization schemes; 25\% of clusters are Bernoulli randomized (i.e., treatment assignment is randomly assigned at the listing level), whereas the remaining 75\% of clusters are cluster randomized (i.e., treatment assignment is randomly assigned at the cluster level). Both of the meta-experiments we present are related to pricing on Airbnb. We focus on pricing-related treatment interventions for two reasons. First, it is crucial for both hosts and the platform intermediary to understand the price elasticity of Airbnb guests; hosts set the price of their listings, while Airbnb recommends prices to hosts and sets platform fees. Second, TATE estimates for pricing-related experiments are likely to be affected by interference bias, since hosts observe other hosts' prices and guests usually consider many listings before choosing a listing to book.

The first meta-experiment measures the effect of a change to Airbnb's platform fee structure. In the treatment group, long-tenured hosts were subject to a platform guest fee increase, while the platform guest fee for less tenured hosts remained unchanged. In the control group, long-tenured hosts were subject to a platform guest fee \textit{decrease}, while the platform guest fee for less tenured hosts remained unchanged. Results from the Bernoulli randomized meta-treatment arm suggest that the treatment led to a statistically significant loss of 0.207 bookings per listing over the course of the experiment.\footnote{To avoid disclosing raw numbers, all raw booking, nights booked, and gross guest spend values are multiplied by a constant.} However, a joint analysis of the entire meta-experimental sample finds that there is a statistically significant difference between the TATE estimates obtained in the two meta-treatment arms. We estimate that 32.60\% of the Bernoulli TATE estimate on bookings is attributable to interference bias. While not statistically significant, we also report results that suggest that interference bias is more severe in markets that are demand constrained than in markets that are supply constrained.

Results from the fee meta-experiment establish the existence of interference bias in online marketplaces, and the efficacy of cluster randomization in reducing that bias. However, the guest platform fee treatment intervention is one that affects all hosts on Airbnb. Often, online marketplace designers are interested in the effect of behavioral nudges, which only cause a change in the behavior of some users. These experiments are typically analyzed with intention-to-treat (ITT) analysis. To test for interference bias in an experiment that requires ITT analysis, we conduct a second meta-experiment that measures the effect of a proposed update to the algorithm underlying Airbnb's price suggestions for hosts. On average, the treatment \textit{increased} the prices suggested to hosts. Results from the Bernoulli randomized meta-treatment arm suggest that the treatment led to a statistically significant loss of 0.106 bookings per listing over the course of the experiment. In the cluster randomized meta-treatment arm, this treatment effect disappears; the point estimate is smaller in magnitude, and not statistically significant. However, a joint analysis of the entire meta-experimental sample fails to detect a statistically significant difference between the two sets of treatment effect estimates. Post-hoc power analysis reveals that the meta-experiment is underpowered to detect interference bias that is not extremely severe in magnitude. Although not statistically significant, our point estimates suggest that in the Bernoulli randomized pricing experiment, 54.16\% of the observed treatment effect is due to interference bias. This result highlights the difficulty of detecting interference bias when a given treatment intervention only affects some users, even if the magnitude of that bias is potentially large.

While previous research has focused on quantifying the magnitude of interference bias through simulation \citep{fradkin.2015, holtz.2018} or post-hoc analysis \citep{blake.2014}, this work is among the first empirical papers to focus on reducing interference bias in a marketplace experiment through experiment design. The experiment design techniques we employ are strongly influenced by the network experimentation literature \citep{eckles.2017, ugander.2013, saveski.2017}, and future extensions of our work might focus on adopting analysis-based approaches to reducing interference bias in network experiments \citep{athey.2018, aronow.2012, eckles.2017, chin.2018} to an online marketplace setting. Future work might also focus on how to best cluster items or sellers in a marketplace. Clustering items or sellers in an online marketplace is difficult, as there is often no explicit network structure indicating which items are likely to substitute or complement for one another,\footnote{When part of an online market's design, recommendation networks \citep{oestricher.2012b, oestricher.2012a} do provide an explicit product network.} and measuring cross-price elasticities in markets with millions of heterogeneous goods is difficult.

The rest of this paper proceeds as follows. In Section \ref{sec:literature}, we review the related literature. In Section \ref{sec:setting}, we describe in greater detail the features of Airbnb's platform that are relevant to the two meta-experiments presented in this paper. Our meta-experiment design is described in Section \ref{sec:experiment_design}. We present results from the fee experiment in Section \ref{sec:fee_experiment}, and results from the pricing algorithm experiment in Section \ref{sec:algorithm_experiment}. Finally, we discuss our findings and future extensions in Section \ref{sec:discussion}.

\section{Related Literature} \label{sec:literature}

The research in this paper connects to three bodies of academic literature: one on interference bias in online marketplace experiments, one on experimentation in networks, and one on pricing-related online marketplace interventions.

Our work is most closely related to recent research that has shown that naive marketplace experimentation can yield total average treatment effect estimates that are overstated by up to 100\% \citep{blake.2014, fradkin.2015, holtz.2018}. \cite{blake.2014} arrive at this conclusion through post-hoc analysis of an experiment conducted on eBay, while \cite{fradkin.2015} finds evidence for interference bias through a simulation of Airbnb's marketplace that has been calibrated using search and transaction data from the firm. Finally, \cite{holtz.2018} also shows through a simple simulation of marketplace experiments on Airbnb that naive marketplace experiments are biased due to interference, and that the magnitude of this bias can be reduced through experiment design and analysis techniques.

Bias in total average treatment estimates due to statistical interference is not a problem unique to online marketplace experiments. In fact, there has been substantial research on experiment design and analysis techniques that provide unbiased TATE estimators in settings where the stable unit treatment value assumption \citep{rubin.1974} is violated.\footnote{SUTVA is sometimes alternatively referred to as the `no interference' assumption \citep{cox.1958}.} SUTVA assumes that the potential outcomes of a given unit of analysis are independent of the treatment assignments other units receive. However, in many settings (e.g., networks, marketplaces) SUTVA is unlikely to hold. When SUTVA is violated, the TATE estimated from a Bernoulli randomized experiment can differ substantially from the actual TATE (i.e., the average effect of the treatment under the counterfactual that every unit is treated). Network science researchers have developed experiment designs \citep{ugander.2013, eckles.2017} and treatment effect estimators \citep{aronow.2012, chin.2018} that eliminate or reduce bias due to SUTVA violations arising from network interference. 

\cite{ugander.2013} propose graph cluster randomization (GCR) as an experiment design for reducing interference bias in networked experiments. In GCR, a network is first clustered, then randomized at the \textit{cluster}-level. This can greatly reduce the probability that any ego's experimental treatment assignment is different from the treatment assignment of its alters. This will reduce the extent to which statistical interference affects experimental TATE estimates. Through simulations, \cite{eckles.2017} show that GCR can be effective in reducing interference bias in networked experiments, even when the network does not satisfy the strict requirements requirements outlined in \cite{ugander.2013}. One drawback of assigning treatment at the cluster-level is that most treatment effect estimators will provide less statistical power than they would have under a Bernoulli randomized design. However, techniques such as regression adjustment and pre-stratification \citep{moore.2012} can be used in tandem with GCR to mitigate the loss of statistical power. Graph cluster randomization can also be used to test whether or not interference bias affects the TATE estimates obtained from a given experiment. \cite{saveski.2017} conduct a ``Meta-experiment" on LinkedIn, which randomizes over two experiment designs (Bernoulli randomization and cluster randomization). By comparing the treatment effect estimates obtained in each meta-treatment arm, they are able to test for the existence of network interference for any experiment conducted on LinkedIn. 

Finally, our work also connects to the literature on pricing-related online marketplace interventions. A number of recent empirical papers measure the effects of pricing-related interventions on online platforms \citep{dube.2017, filippas.2019}. Airbnb itself uses a customized regression model to provide pricing recommendation to hosts \citep{ifrach.2016, ye.2018}. It is crucial for both platform intermediaries and platform sellers to understand the price elasticity of their customers; sellers would like to price effectively, whereas intermediaries would like to implement effective fee structures and pricing-related market mechanisms. However, TATE estimates obtained through naive experimental tests of pricing-related interventions will likely yield biased estimates of price elasticity, since marketplace sellers compete with one another, and observe each others' pricing decisions. 

This paper builds on prior research by adapting experiment design techniques from the networks literature \citep{ugander.2013, eckles.2017, holtz.2018} and conducting meta-experiments \citep{saveski.2017} in an online marketplace to test for the existence of interference bias. Developing methods for obtaining accurate TATE estimates in online marketplace settings is increasingly important as both researchers and practitioners continue to explore novel pricing-related interventions \citep{dube.2017, filippas.2019} in online marketplace settings.

\section{Setting} \label{sec:setting}

Airbnb is an online marketplace for accommodations and experiences. More than five million listings appear on Airbnb, and since the company's founding in 2008, over 400 million guest arrivals have occurred on the platform. On average, over two million people are staying in Airbnb listings on a given night \citep{airbnb.press}. 

\subsection{Platform Guest Fees}

\begin{figure}[htpb]
  \centering
    \includegraphics[width=\textwidth]{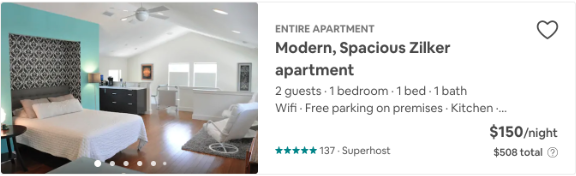}
      \caption{A typical search result on Airbnb. For this search result, the guest platform fee is included in the total price of \$508.}
      \label{fig:p2}
\end{figure}

\begin{figure}[htpb]
  \centering
    \includegraphics[width=\textwidth]{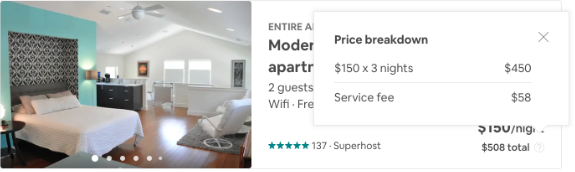}
      \caption{The price breakdown tooltip for a typical search result on Airbnb. In this tooltip, the guest platform fee (listed here as a service fee of \$58) is broken out from the nightly price.}
      \label{fig:p2_tooltip}
\end{figure}

\begin{figure}[htpb]
  \centering
    \includegraphics[]{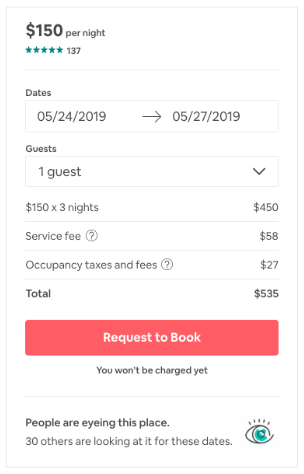}
      \caption{The section of the Airbnb product detail page that provides a full pricing breakdown for would-be guests. In this pricing breakdown, the guest platform fee (listed here as a service fee) is \$58.}
      \label{fig:pdp}
\end{figure}

Airbnb earns revenue by collecting fees from guests and hosts for every transaction that occurs on the platform. In order to set fees optimally, it is crucial for the platform to understand guest price elasticity. Airbnb's fees for guests are visible in three different locations throughout the booking process. First, guest platform fees are included in the total price shown to guests when a listing appears in search. Figure \ref{fig:p2} shows a typical Airbnb search result. Second, if a guest opens a tooltip on any search result, they are shown a price breakdown that separates the listing's nightly price and the guest platform fee. Figure \ref{fig:p2_tooltip} shows this tooltip. Finally, when viewing a listing's product detail page, a detailed pricing breakdown (including fees) is displayed next to the ``Request to Book" button. Figure \ref{fig:pdp} shows this price breakdown.

\subsection{Price Tips \& Smart Pricing}

Since the summer of 2015, Airbnb has provided tools to help hosts price more effectively. In June 2015, Airbnb launched ``Price Tips," a feature that provides dynamic pricing suggestions for hosts \citep{airbnb.pricetips}. In November 2015, Airbnb launched ``Smart Pricing," a tool that automatically updates hosts' prices subject to a set of constraints determined by the host \citep{airbnb.smartpricing}. Both ``Price Tips" and ``Smart Pricing" present recommendations from the same machine learning model, which incorporates local supply and demand features to provide dynamic pricing suggestions to hosts \citep{ifrach.2016, ye.2018}. We refer the reader to \cite{ye.2018} for a more detailed description of the pricing algorithm itself. Importantly, Airbnb's pricing suggestions attempt to maximize each host's individual objectives, rather than playing the role of a central planner.  

\begin{figure}[htpb]
  \centering
    \includegraphics[width=\textwidth]{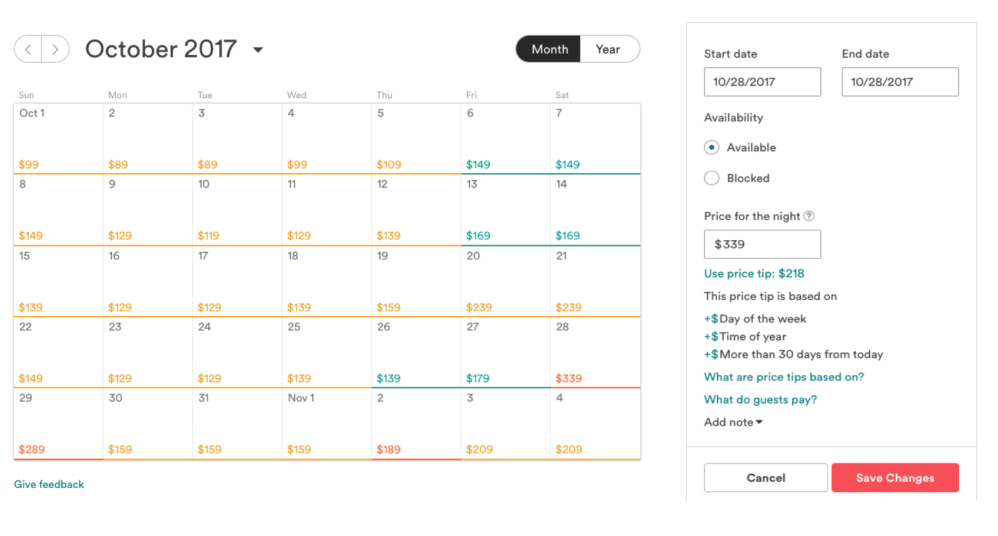}
      \caption{A screenshot of the ``Price tips" UI. ``Price tips" color codes the nights on a host's calendar based on the pricing model's estimated probability that a given night will be booked. If a host selects a given calendar night, the host is shown the model's suggested price. Airbnb also presents explanations for why it is recommending a particular price (e.g., ``Time of year," ``More than 30 days from today"). In order for a given host to fully adopt Airbnb's recommended prices with the ``Price tips" product, the host is required to visit Airbnb every day, review Airbnb's recommendations, and accept them. Image from \cite{ye.2018}.}
      \label{fig:price_tips}
\end{figure}

\begin{figure}[htpb]
  \centering
    \includegraphics[width=\textwidth]{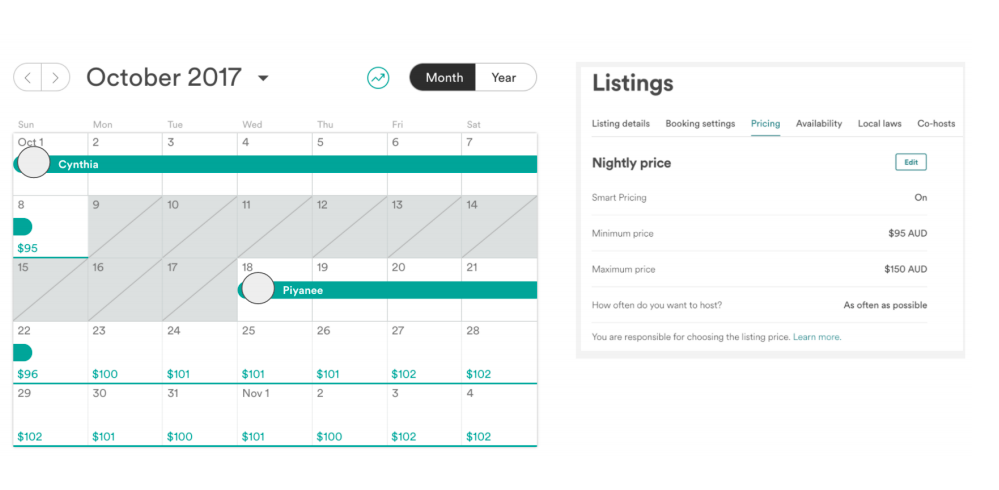}
      \caption{A screenshot of the ``Smart pricing" UI. When setting up``Smart Pricing," hosts provide a minimum and maximum price. After ``Smart Pricing" is turned on, hosts automatically adopt Airbnb's recommended price if it is between the host's minimum and maximum price. If Airbnb's recommendation is higher than the host's upper threshold, the price is set to the upper threshold. If Airbnb's recommendation is lower than the host's lower threshold, the price is set to the lower threshold. A screenshot of the ``Smart Pricing" UI is shown in Figure \ref{fig:smart_pricing}. Image from \cite{ye.2018}.}
      \label{fig:smart_pricing}
\end{figure}

``Price tips" color codes nights on a host's calendar based on the estimated probability that a given night will be booked given the current price, and suggests an ``optimal" price for each night. Importantly, ``Price tips" requires hosts to manually accept prices in order to comply with the algorithm's suggestions recommended through the ``Price tips" product. A screenshot of the ``Price tips" UI is shown in Figure \ref{fig:price_tips}. ``Smart pricing" was introduced to make it easier for hosts to adopt Airbnb's pricing recommendations en masse. Once ``Smart pricing" is turned on, hosts automatically adopt Airbnb's recommended price, subject to constraints provided by the host. A screenshot of the ``Smart Pricing" UI is shown in Figure \ref{fig:smart_pricing}.

\section{Experiment Motivation \& Design} \label{sec:experiment_design}

It is crucial for an online marketplace intermediary, such as Airbnb, to understand the price elasticity of its customers. This enables the firm to implement optimal pricing-related market mechanisms, such as fee structures and seller pricing suggestions. Understanding customer price elasticities can also be beneficial to sellers, who set their own prices. If the business outcomes of all Airbnb listing were independent, the firm could take an atheoretic approach to estimating price elasticity by running a randomized controlled trial, or A/B test, in which the prices of some listings were exogenously increased or decreased. However, as described in \cite{holtz.2018}, host- or listing-level experiments on Airbnb violate SUTVA due to the inherent interconnectedness of online marketplaces. 

There are a number mechanisms that can lead to SUTVA violations on Airbnb. For one, if some hosts lower (raise) their prices, they may increase (decrease) demand for their listings, and, consequently, decrease (increase) demand for their competitors' listings. Furthermore, host pricing decisions may exhibit viral properties; a host may observe their competitor's pricing behavior, and copy it. Finally, Airbnb listings in a given market can also serve as complements to each other. For instance, guests may describe their positive (negative) experience with a given listing to their peers, which could increase (decrease) demand for similar listings.

Adapting experiment design and analysis techniques from the network experimentation literature, as proposed by \cite{holtz.2018}, is one avenue for reducing interference bias in online marketplace pricing experiments. However, none of the techniques put forward by \cite{holtz.2018} have been used yet to design or analyze an online marketplace experiment. As a first step toward empirically confirming the existence of interference bias in online marketplace experiment TATE estimates, and measuring the extent to which cluster randomization, an experiment design technique, can reduce that bias, we conduct pricing-related meta-experiments \citep{saveski.2017} on Airbnb. Quantifying the magnitude of interference bias, as well as the extent to which cluster randomization can reduce that bias, is useful for two reasons. First, even if interference bias is a theoretical concern, it may not be a practical one; statistical bias in TATE estimates due to interference may be small. Second, even if interference bias is large, cluster randomization may not be an effective tool to reduce that bias. If this were the case, cluster randomization would not be a worthwhile undertaking for firms; cluster randomization results in reduced statistical power relative to Bernoulli randomization, and is also more logistically complicated for firms to implement (both because of the need to identify relevant clusters, and because most corporate A/B testing tools do not support cluster randomization).

In each meta-experiment, Airbnb listings are arranged into clusters. Each of these clusters is then assigned to one of two meta-treatment arms: Bernoulli randomization, or cluster randomization. Within the Bernoulli-randomized meta-treatment arm, treatment is randomly assigned at the listing level. Within the cluster-randomized meta-treatment, treatment is randomly assigned at the cluster level. By jointly analyzing the data from both meta-treatment arms, we are able to measure whether there is a statistically significant difference between the TATEs measured separately in each meta-treatment arm.

\subsection{Treatment Assignment Mechanism}

In this subsection, we describe the procedure used to arrange Airbnb listings into clusters, and then subsequently determine a given listing's meta-treatment assignment and treatment assignment.

\subsubsection{Clusters of Airbnb Listings}

To perform cluster randomization, it is first necessary to arrange all of Airbnb's listings into mutually exclusive clusters. Previous work \citep{holtz.2018} has proposed creating a network of listings that substitute for or complement one another, and then clustering that network with any of a number of graph clustering algorithms (e.g., Louvain clustering \citep{blondel.2008}). In this subsection, we outline a different approach to clustering, which we use to generate our listing clusters. We first generate a dense, 16-dimensional demand embedding for each listing, and then cluster listings based on their location in that 16-dimensional space. Our method for generating Airbnb listing embeddings is similar to that described in \cite{grbovic.2018}.

Our embeddings are trained on data consisting of sequences of listings that individual users view in the same search session. If, for instance, a user viewed listings $L_{A}$, $L_{B}$, and $L_{C}$ in one search session, this would generate the sequence:

\begin{equation}
<L_{A}, L_{B}, L_{C}>.
\end{equation} 

\noindent We use a word2vec-like architecture \citep{mikolov.2013b} to estimate a skip-gram model \citep{mikolov.2013a} on this data. Given $S$ sequences of listings, the skip-gram model attempts to maximize the objective function

\begin{equation}
J = \textrm{max}_{W, V} \sum_{s \in S} \frac{1}{|s|} \sum_{i=1}^{|s|} \sum_{-k \leq j \leq k, \, k \neq 0} \log p \left ( L_{i+j} | L_i \right ),
\end{equation}

\noindent where $k$ is the size of a fixed moving window over the listings in a session, $W$ and $V$ are weight matrices in the word2vec architecture, and $p(L_{i+j} | L_i)$ is the hierarchical Softmax approximation to the regular softmax expression.

The objective function above is augmented by including listing-level attributes (e.g., a listing's market) in the search session sequences. The model is then trained using a market-level negative sampling approach. This generates a 16-dimensional vector representation for each Airbnb listing.

Once listing embeddings are estimated using the aforementioned approach, a recursive partitioning tree \citep{kang.2016} is used to arrange the Airbnb listings into clusters. The algorithm starts from a single cluster containing all listings, and then recursively bisects clusters into two sub-clusters. The algorithm stops bisecting sub-clusters when the tree reaches a depth of 20, or when a new sub-cluster will contain less than 20 listings. Listings can then be assigned to clusters of arbitrary sizes by assigning them to the smallest sub-cluster to which they belong that has at least some threshold number of listings. For the algorithmic pricing meta-experiment, we set this threshold at 250 listings, whereas for the fee meta-experiment, we set this threshold at 1,000 listings.\footnote{In choosing cluster sizes, we are attempting to balance two objectives: creating clusters that capture listings likely to interfere with one another, and designing an experiment with sufficient statistical power. Since ex ante, we expected the fee treatment intervention to have a larger effect, we chose larger clusters for that meta-experiment. For more details on the process used to determine cluster size, see Appendix \ref{sec:cluster_size_selection}.} Figure \ref{fig:cluster_examples} depicts example clusters generated using this method in the Bay Area.

\subsubsection{Pre-stratification \& Treatment Assignment}

Once listings have been assigned to clusters, those clusters are given meta-treatment assignments and, based on those cluster-level meta-treatment assignments, listings are assigned listing-level treatment assignments. 

To gain statistical power (particularly in the cluster-randomized meta-treatment arm), we group clusters into strata using a multivariate blocking procedure \citep{moore.2012}. As a first step, we collected pre-treatment listing-level data.\footnote{For the fee meta-experiment, pre-treatment data was collected from January 16, 2019 to February 17, 2019. For the pricing algorithm experiment, pre-treatment data was collected from August 1, 2018 to September 25, 2018.} We then aggregate data at the cluster level, and for each cluster calculate over the pre-treatment period the average number of nights booked per listing, the average number of bookings per listing, the average booking value per listing, and the number of experiment-eligible listings in the cluster.\footnote{Our experiment excludes listings in a long-term experiment holdout group, as well as listings in Airbnb's ``Plus" tier.}\footnote{For the algorithmic pricing experiment, we also calculate the percentage of listings accepting at least one price tip during the pre-treatment period, and the percentage of listings with ``Smart Pricing" enabled at the end of the pre-treatment period.} After centering and scaling each of these metrics, we calculate the Mahalanobis distance between each pair of clusters. Finally, we use an optimal-greedy algorithm to arrange clusters into strata of size $n=8$. At each step, the optimal-greedy algorithm finds the smallest ``available" distance between two clusters\footnote{A distance is ``available" if that pair of clusters has not been used in a previous step.}, and assigns the two corresponding clusters to the same stratum. 

Within each stratum, two clusters are assigned to the meta-control via complete random assignment. The remaining six clusters are assigned to the meta-treatment. Within the meta-control arm, Bernoulli randomization is used to assign 50\% of listings to the treatment and 50\% of listings to the control. Within the meta-treatment arm, three of the six clusters are assigned the treatment via complete random assignment. The remaining three clusters are assigned the control. Each listing in a meta-treatment cluster is assigned the treatment assignment corresponding to its cluster.

\section{Fee Meta-experiment} \label{sec:fee_experiment}

\subsection{Description}

The fee meta-experiment ran from March 16, 2019 to March 21, 2019 on a population of 4,578,028 listings. Of those listings, 1,146,537 were assigned to the Bernoulli-randomized meta-treatment arm, and the remaining 3,431,491 were assigned to the cluster-randomized meta-treatment arm. Within the Bernoulli-randomized meta-treatment arm, 573,346 were assigned to the treatment and 573,191 listings were assigned to the control. Within the cluster-randomized meta-treatment arm, 2,982 clusters were assigned to the treatment and 2,982 clusters were assigned to the control, resulting in 1,720,147 listings assigned to the treatment and 1,711,344 listings assigned to the control. In total, across both meta-treatment arms, 2,293,493 listings were assigned to the treatment, and 2,284,535 were assigned to the control. 

Figure \ref{fig:ecdf_fee_experiment} shows the empirical CDFs for pre-treatment bookings, nights booked, and booking value across all four meta-treatment / treatment groups.\footnote{To avoid disclosing raw numbers, x-axis values are multiplied by a constant.} For each of these pre-treatment outcomes, the empirical CDFs are visually quite similar. Kolmogorov-Smirnoff tests comparing the distributions for each pre-treatment outcome in the meta-treatment and meta-control fail to reject the null of equal distributions for bookings ($p$ = 0.069) and nights booked ($p$ = 0.647), but do reject the null for booking value ($p$ = 0.021). Kolmogorov-Smirnoff tests comparing the distributions for each pre-treatment outcome in the treatment and control groups of the meta-control arm fail to reject the null of equal distributions for bookings ($p$ = 1.000), nights booked ($p$ = 0.627), and booking value ($p$ = 0.883).  Kolmogorov-Smirnoff tests comparing the distributions for each pre-treatment outcome in the treatment and control groups of the meta-treatment arm fail to reject the null of equal distributions for bookings ($p$ = 0.295), but \textit{do} reject the null of equal distributions for nights booked ($p$ = 0.012) and booking value ($p$ =0.001). We believe that some KS tests fail because cluster randomization is a higher variance randomization procedure, due to both the smaller number of units over which randomization occurs and the correlation of pre-treatment outcomes among listings in the same cluster. Furthermore, booking value is a higher variance outcome than nights booked, which itself is a higher variance outcome than bookings.

In the fee meta-experiment, listings in the treatment had their fees increased if they were long-tenured listings (i.e., if they had been on the platform as of a certain cutoff date). Listings in the control had their fees \textit{decreased} if they were long-tenured listings. In both treatment arms, less tenured listings (i.e., those created after the cutoff date) did not have their fees changed.\footnote{Due to confidentiality concerns on behalf Airbnb, we are unable to disclose the exact magnitude of the fee changes in this experiment, nor are we able to disclose the cutoff date.} Conceptually, one can think of the treatment and control conditions of this meta-experiment as comparing the effect of two different fee-based incentive programs Airbnb might run. In the treatment group, new listings have lower fees (which could drive business to newer listings), whereas in the control, older listings have lower fees (which could reward long-time Airbnb hosts and reduce churn). After the conclusion of the fee meta-experiment, a ``reversal experiment" was run from April 15, 2019 to April 22, 2019. In the reversal experiment, listings that had been assigned the treatment condition in the meta-experiment were assigned the control, and vice-versa. The purpose of the reversal experiment was to mitigate any negative impact of the meta-experiment on Airbnb hosts.

\subsection{Results}

In this section, we present results from the fee meta-experiment. We focus on a single outcome metric, bookings per listing, but the results for two alternative outcome metrics, nights booked per listing and gross guest spend per listing, are qualitatively similar and can be found in Appendix \ref{sec:nb_and_bv}.\footnote{To avoid disclosing raw numbers, all raw booking, nights booked, and gross guest spend values are multiplied by a constant.} Since, relative to the control, the treatment \textit{increased} fees, we expect the TATE on bookings per listing to be negative. 

We first present the results from separately analyzing the Bernoulli randomized arm of the meta-experiment and the cluster randomized arm of the meta-experiment. While the Bernoulli randomized arm will have ample statistical power, we expect its TATE estimate to suffer from interference bias. On the other hand, analysis of the cluster randomized arm should provide a less biased estimate of the TATE, since the amount of marketplace interference will be reduced, but will also have less statistical power. Simply comparing the point estimates obtained independently from the two meta-treatment arms is not sufficient to rigorously measure interference bias. In order to do so, we proceed to jointly analyze both the Bernoulli randomized and cluster randomized meta-treatment arms. Finally, we investigate the extent to which our results are contingent on how supply- or demand-constrained a given Airbnb market is.

\subsubsection{Bernoulli \& Cluster Randomized Results}

We analyze both the Bernoulli randomized and cluster randomized meta-treatment arms separately by estimating the following model,

\begin{equation} \label{eq:main_experiment_spec}
Y_i = \alpha + \beta T_i + \sum_l \gamma_l \mathbbm{1}(B_i = l) + \delta X_i + \epsilon_i
\end{equation}

\noindent on listing-level data, where $Y_i$ is the outcome of interest, $T_i$ is the treatment assignment for listing $i$, $B_i$ is a variable indicating which stratum listing $i$'s cluster of belongs to, $X_i$ is a vector consisting of listing $i$'s pre-treatment bookings, nights booked, booking value, and gross guest spend, and $\epsilon_i$ is an error term.\footnote{Data from the cluster randomized meta-treatment arm can also be analyzed by first aggregating the data at the cluster level and then estimating a weighted version of Equation \ref{eq:main_experiment_spec}. We present this analysis in Appendix \ref{sec:cluster_agg_analysis}. This analysis results in estimates that are nearly identical to those obtained by analyzing the experiment with listing-level data.}  For all analyses, we cluster standard errors at the Airbnb listing cluster-level.

\begin{figure}[htpb]
  \centering
    \includegraphics[width=\textwidth]{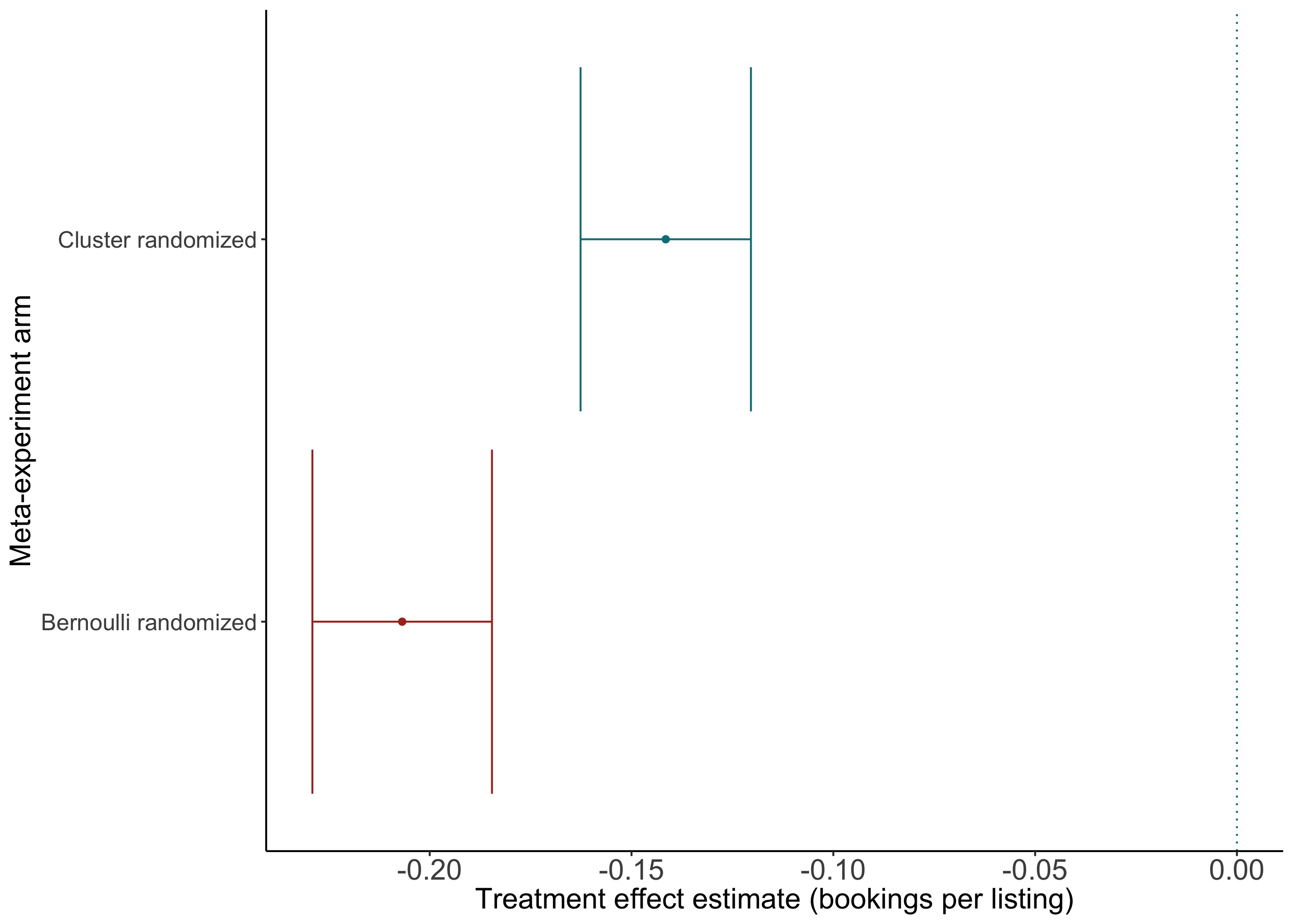}
      \caption{Total average treatment effect estimates for the fee experiment, estimated separately in the Bernoulli randomized meta-treatment arm and the cluster randomized meta treatment arm. Error bars represent 95\% confidence intervals. The dotted blue line corresponds to a treatment effect of 0 bookings per listing.}
      \label{fig:arm_level_results_fees_bookings}
\end{figure}

Table \ref{tab:experiment_analysis_fees_bookings} shows the TATE estimate for bookings per listing in both the Bernoulli randomized and cluster randomized meta-treatment arms. In the Bernoulli randomized meta-treatment arm, the TATE is -0.207 bookings per listing, whereas in the cluster randomized meta-treatment arm, the TATE is -0.142 bookings per listing. Both of these TATE estimates are statistically significant at the 95\% confidence level. Figure \ref{fig:arm_level_results_fees_bookings} shows the estimated TATE in both meta-treatment arms, along with the corresponding 95\% confidence intervals.

Although the TATE estimates obtained from the two meta-experiment arms are different, it is not clear when analyzing the two meta-experiment arms separately whether or not there is a statistically significant difference between the two estimates. By extension, it is still unclear whether or not the Bernoulli TATE estimate suffers from interference bias and/or if cluster randomization helps to mitigate this bias. In order to rigorously test for a difference, it is necessary to jointly analyze both meta-treatment arms simultaneously.

\subsubsection{Joint Analysis}

In order to determine with statistical rigor whether the two meta-treatment arms yield different treatment effect results, we estimate the model,

\begin{equation} \label{eq:meta_experiment_spec}
Y_i = \alpha + (\beta + \nu M_i) T_i + \xi M_i + \sum_l \gamma_l \mathbbm{1}(B_i = l) + \delta X_i + \epsilon_i,
\end{equation}

\noindent where $Y_i$ is the outcome of interest, $M_i$ is a binary variable set to 1 when listing $i$ is in the Bernoulli meta-treatment arm and 0 when $i$ is in the cluster-randomized meta-treatment arm, $T_i$ is a binary variable set to 1 when listing $i$ is exposed to the treatment, $B_i$ is a variable indicating the stratum of clusters to which listing $i$ belongs, $X_i$ is a vector consisting of listing $i$'s pre-treatment variables, and $\epsilon_i$ is the error term. Once again, we cluster standard errors at the Airbnb listing cluster-level.

In the above model, $\beta$ measures the ``true" effect of the treatment,\footnote{Even when using cluster randomization, TATE estimates may be biased, since clusters do an imperfect job of capturing listings that complement and substitute for one another. Furthermore, interference may extend beyond a given listing's immediate substitutes or complements.} and $\nu$ measures the difference between the effect of the treatment in the Bernoulli arm and the effect of the treatment in the cluster randomized arm. In other words, $\nu$ should measure the extent to which cluster randomization reduces interference bias. $\xi$ measures any baseline difference between listings in the Bernoulli-randomized arm of the meta-experiment and listings in the cluster-randomized arm of the meta-experiment. Since clusters were assigned to meta-treatment arms using the random assignment procedure described in Section \ref{sec:experiment_design}, we expect $\xi$ to be zero. However, it is possible that imbalances between listings in the two meta-treatment arms persist even after random assignment.

Table \ref{tab:meta_experiment_analysis_fees_bookings_only} shows the results from estimating Equation \ref{eq:meta_experiment_spec} for the fee meta-experiment using listing level data.\footnote{Joint meta-experiment data can also be analyzed using a weighted combination of individual listing-level data from the Bernoulli randomized meta-treatment arm and aggregated cluster-level data from the cluster randomized meta-treatment arm. This analysis results in estimates that are nearly identical to those obtained using listing-level data from both meta-treatment arms. We present this analysis in Appendix \ref{sec:mixed_analysis}.} Figure  \ref{fig:meta_analysis_plot_fees_bookings_only} displays our point estimate for each parameter in Equation \ref{eq:meta_experiment_spec}, along with 95\% confidence intervals. We estimate that the ``true" TATE is -0.139 bookings per listing, whereas -0.067 bookings per listing of the TATE measured in the Bernoulli randomized meta-treatment arm is due to interference bias. In other words, we estimate that 32.60\% ($\pm12.93\%$) of the TATE estimate achieved through a Bernoulli randomized experiment is due to interference bias, and is eliminated by instead running a cluster randomized experiment. 

\subsubsection{The Moderating Effect of Supply and Demand Constrainedness}

Given that interference bias arises in part due to substitution and complementarity between Airbnb listings, one might expect that the extent to which interference causes bias in the Bernoulli randomized TATE estimate depends on the conditions in a given Airbnb market. For instance, interference bias may be \textit{more severe} in markets that are demand constrained, and \textit{less severe} in markets that are supply constrained.

In order to test this hypothesis, we re-estimate Equation \ref{eq:meta_experiment_spec} for subsets of Airbnb listings that are located in particularly supply constrained or demand constrained markets. Airbnb calculates a supply elasticity index and demand elasticity index for all markets that are above some threshold size using a Cobb-Douglas matching model a la \cite{fradkin.2015}. Of the markets for which these indices are calculated, we keep data for listings that are in markets larger than the median market (computed at the listing level). We then define a listing as being in a \textit{supply constrained} market if its market's supply elasticity index is above the 75th quantile of supply elasticity indices (computed at the listing level), and define a listing as being in a \textit{demand constrained} market if its market's demand elasticity index is above the 75th quantile of demand elasticity indices (computed at the listing level). 

Column 1 of Table \ref{tab:meta_experiment_analysis_fees_supply_demand_constrained} shows our results for supply constrained listings, and Column 2 of Table \ref{tab:meta_experiment_analysis_fees_supply_demand_constrained} shows our results for demand constrained listings. Neither joint analysis is able to detect interference bias with statistical significance. However, if we take our non-statistically significant point estimates as given, our results do suggest that interference bias accounts for 15.09\% of the Bernoulli TATE estimate in demand constrained markets, whereas interference bias actually \textit{reduces} the magnitude of the Bernoulli TATE estimate by 27.41\% in supply constrained markets. We interpret this as weak evidence that interference bias is more likely to lead to inflated TATE estimates in demand constrained markets than supply constrained markets, although further research should be conducted to better understand this relationship.

\section{Algorithmic Pricing Experiment} \label{sec:algorithm_experiment}

The fee meta-experiment results prove that interference bias can have large effects on the accuracy of TATE estimates for online marketplace experiments, and that cluster randomization can help to minimize interference bias. However, the treatment intervention in the fee meta-experiment, a uniform fee change to a well-defined set of Airbnb listings, is only one of the many types of intervention that may be of interest to practitioners. In fact, many of the interventions that online marketplace designers may want to test are behavioral nudges, which require ITT analysis. In the Airbnb context, one such intervention is a change to Airbnb's algorithmic pricing suggestions for hosts. 

Previous academic research suggests that smaller firms (e.g., Airbnb hosts) often behave ``behaviorally" and act sub-optimally when making managerial decisions \citep{kremer.2019}, including pricing decisions \citep{dellavigna.2017}. Airbnb uses a machine learning model \citep{ifrach.2016, ye.2018} to suggest prices to hosts and help them achieve their business goals. Field experiments have shown that managerial training can lead to increased performance for small firms \citep{bloom.2013, bruhn.2018}, suggesting that Airbnb's algorithmic pricing suggestions can change the behavior of hosts and affect their business outcomes.

When Airbnb tests a new iteration of its pricing algorithm, not all hosts are directly affected. Some hosts do not use Airbnb's pricing suggestions at all, and hosts who access Airbnb's pricing tips through ``Price Tips" often have low compliance rates due to the manual effort required to follow Airbnb's suggestions. Even those hosts who opt into ``Smart Pricing" may not fully comply with Airbnb's new suggestions, since Airbnb's suggestions are often constrained by business logic imposed by the host. Although Airbnb's pricing algorithm experiments do not directly affect all hosts, ITT analysis is required for two reasons. First, the set of hosts who \textit{do} accept Airbnb's suggestions (and the extent to which they comply with those suggestions) is endogenous. Second, the firm is interested in the overall effect of the intervention, including the rate at which hosts accept a given set of suggestions.

In order to test the efficacy with which cluster randomization mitigates interference bias for interventions that require ITT analysis, we present the results from a second meta-experiment in which the treatment intervention is a change to Airbnb's pricing suggestions.

\subsection{Description}

The algorithmic pricing meta-experiment ran from September 28, 2018 to October 31, 2018 on a population of 4,557,234 listings. Of those listings, 1,139,240 were assigned to the Bernoulli-randomized meta-treatment arm, and the remaining 3,417,994 were assigned to the cluster-randomized meta-treatment arm. Within the Bernoulli-randomized meta-treatment arm, 569,821 listings were assigned to the treatment and 569,419 listings were assigned to the control. Within the Cluster-randomized meta-treatment arm, 11,631 clusters were assigned to the treatment, and 11,631 clusters were assigned to the control, resulting in 1,709,018 listings assigned to the treatment, and 1,708,976 listings assigned to the control. In total, across both meta-treatment arms, 2,278,839 listings were assigned to the treatment, and 2,278,395 listings were assigned to the control. Importantly, the sample size for the algorithmic pricing meta-experiment is approximately equal to the sample size for the fee meta-experiment.

Figure \ref{fig:ecdf_pricing_experiment} shows the empirical CDFs for pre-treatment bookings, nights booked, and booking value across all four meta-treatment / treatment groups.\footnote{To avoid disclosing raw numbers, x-axis values are multiplied by a constant.} For each of these pre-treatment outcomes, the empirical CDFs are visually quite similar. Kolmogorov-Smirnoff tests comparing the distributions for each pre-treatment outcome in the meta-treatment and meta-control fail to reject the null of equal distributions for bookings ($p$ = 0.387), nights booked ($p$ = 0.222), and booking value ($p$ = 0.180). Kolmogorov-Smirnoff tests comparing the distributions for each pre-treatment outcome in the treatment and control groups of the meta-control arm fail to reject the null of equal distributions for bookings ($p$ = 1.000), nights booked ($p$ = 0.888), and booking value ($p$ = 0.752). Kolmogorov-Smirnoff tests comparing the distributions for each pre-treatment outcome in the treatment and control groups of the meta-treatment arm reject the null of equal distributions for bookings ($p$ = 0.021) and nights booked ($p$ = 0.021), but fail to reject the null of equal distributions for booking value ($p$ = 0.847). We believe that some KS tests fail for the same reasons as were outlined when describing the fee meta-experiment.

For listings in the treatment group, the suggested prices surfaced through both ``Price Tips" and ``Smart Pricing" were generated by a new version of Airbnb's pricing algorithm. Relative to the status quo algorithm, the treatment algorithm generally increased prices. For instance, on unconstrained smart pricing nights (e.g., calendar nights in which hosts had opted into smart pricing and the suggested price was not subject to a minimum or maximum price threshold), prices increased by 4\% on average.\footnote{Unconstrained smart pricing nights represent only a fraction of the total calendar nights on Airbnb. As a result, the average price increase across \textit{all} calendar nights is less than 4\%.}

\subsection{Results}

In this section, we present results from the algorithmic pricing experiment. As was true for the fee meta-experiment, we report effects of the treatment on bookings per listing, but found qualitatively similar results for nights booked per listing and gross guest spend per listing, which can be found in Appendix \ref{sec:nb_and_bv}.\footnote{To avoid disclosing raw numbers, all raw booking, nights booked, and gross guest spend values are multiplied by a constant.} Since, on average, the treatment pricing algorithm \text{increased} prices, we expect the TATE on bookings per listing to be negative. We first present the results separately analyzing the Bernoulli randomized arm of the meta-experiment and the cluster randomized arm of the meta-experiment. We then proceed to jointly analyze both meta-treatment arms, in order to test for the existence of interference bias in the Bernoulli randomized experiment's TATE estimate.

\subsubsection{Bernoulli \& Cluster Randomized Results}

We analyze both the Bernoulli randomized and cluster randomized meta-treatment arms separately by estimating equation \ref{eq:main_experiment_spec} on listing-level data.\footnote{As was the case with the fee meta-experiment, we present aggregate-level analysis of the cluster randomized meta-treatment arm in Appendix \ref{sec:cluster_agg_analysis}. The results from this analysis are nearly identical.} As was the case with the fee meta-experiment, standard errors are clustered at the Airbnb listing-cluster level.

\begin{figure}[htpb]
  \centering
    \includegraphics[width=\textwidth]{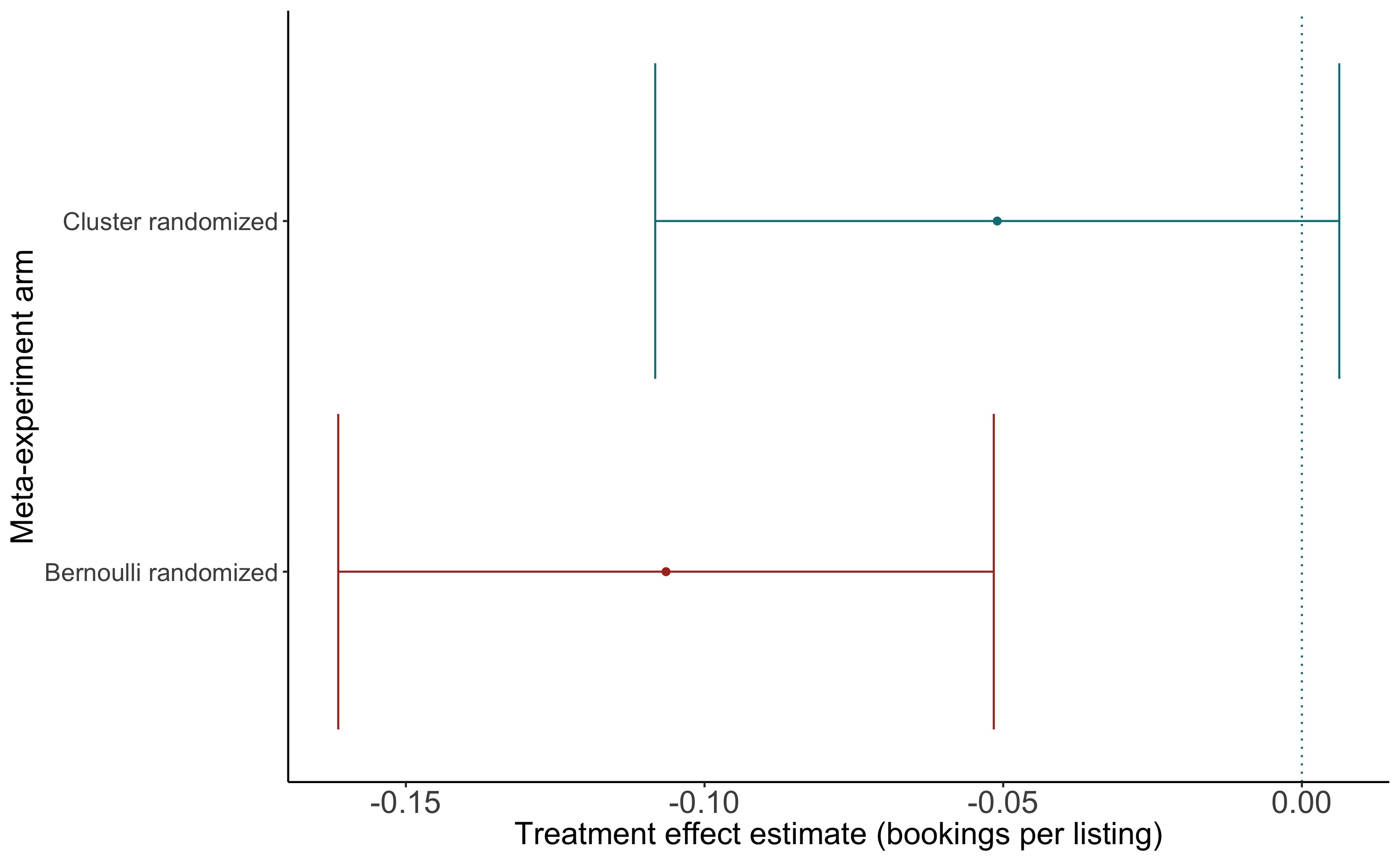}
      \caption{Total average treatment effect estimates for the algorithmic pricing experiment, estimated separately in the Bernoulli randomized meta-treatment arm and the cluster randomized meta treatment arm. Error bars represent 95\% confidence intervals. The dotted blue line corresponds to a treatment effect of 0 bookings per listing.}
      \label{fig:arm_level_results_pricing_bookings}
\end{figure}

Table \ref{tab:experiment_analysis_pricing_bookings} shows the TATE estimate for bookings per listing in both the Bernoulli randomized and cluster randomized meta-treatment arms. In the Bernoulli randomized meta-treatment arm, the TATE is -0.106 bookings per listing, and this result is statistically significant at the 95\% confidence level. In the cluster randomized meta-treatment arm, our point estimate of the TATE is -0.051 bookings per listing, however, this result is not statistically significant at the 95\% confidence level. Figure \ref{fig:arm_level_results_pricing_bookings} shows the estimated TATE in both meta-treatment arms, along with the corresponding 95\% confidence intervals. In order to rigorously test whether or not cluster randomization led to a reduction in interference bias, we proceed to jointly analyze both meta-treatment arms.

\subsubsection{Joint Analysis}

In order to determine whether or not the two meta-treatment arms yield TATE estimates between which there is a statistically significant difference, we once again estimate equation \ref{eq:meta_experiment_spec}.\footnote{For the algorithmic pricing meta-experiment, $X_i$ also includes listing $i$'s smart pricing opt-in status at the outset of the experiment.} As was the case with the fee meta-experiment, standard errors are clustered at the Airbnb listing-cluster level.

\begin{figure}[htpb]
  \centering
    \includegraphics[width=\textwidth]{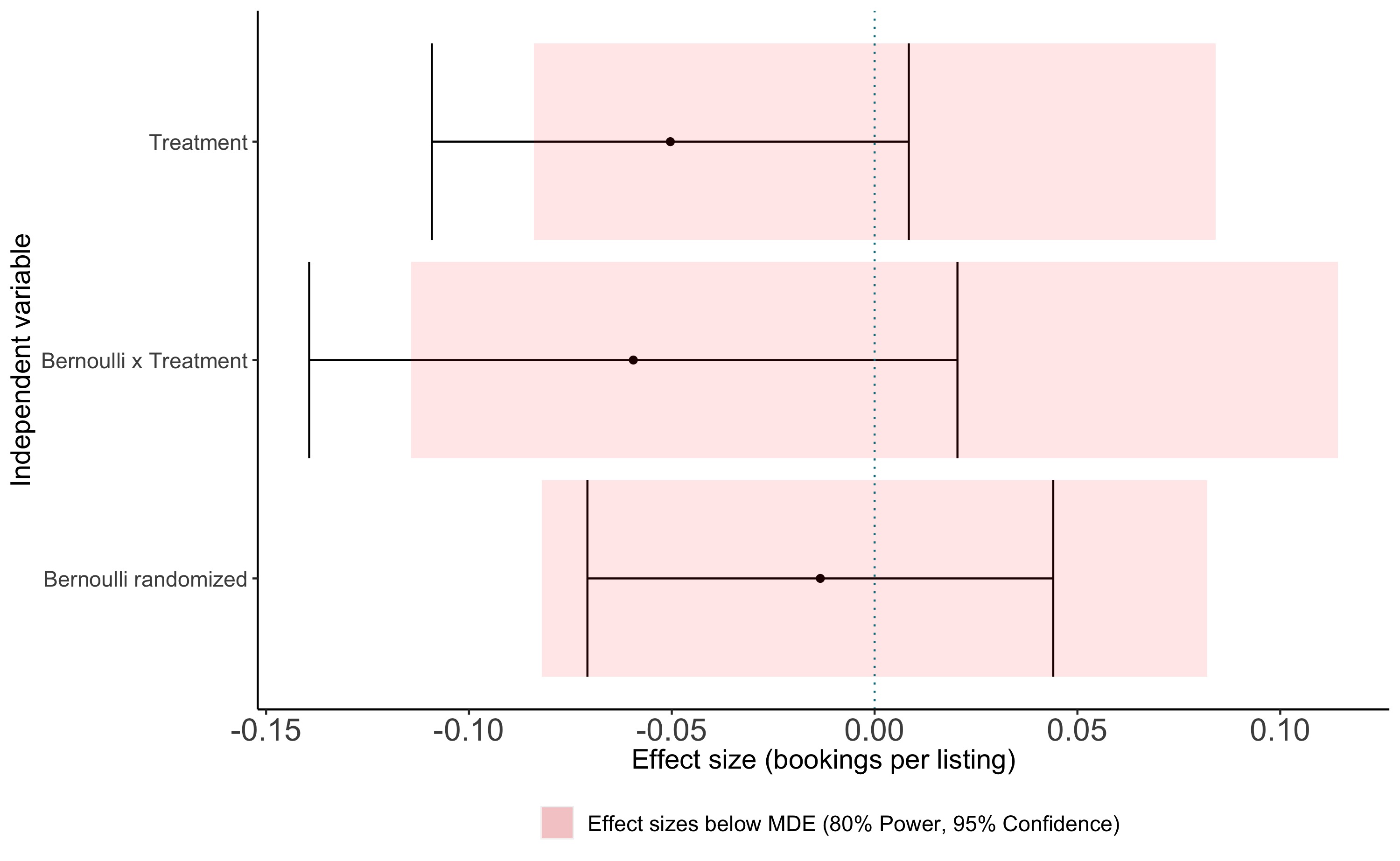}
      \caption{Coefficient estimates for the joint analysis of the algorithmic pricing meta-experiment. Error bars represent 95\% confidence intervals. The dotted blue line correponds to a treatment effect of 0 bookings per listing. The red shaded area corresponds to values that are below the MDE (80\% power, 95\% confidence).}
      \label{fig:meta_analysis_plot_pricing_bookings_only}
\end{figure}

Table \ref{tab:meta_experiment_analysis_pricing_bookings_only} shows our results, and Figure \ref{fig:meta_analysis_plot_pricing_bookings_only} displays our point estimate for each parameter in Equation \ref{eq:meta_experiment_spec}, along with 95\% confidence intervals. Point estimates imply that the ``true" TATE is -0.050 bookings per listing, whereas -0.059 bookings per listing of the TATE measured in the Bernoulli randomized meta-treatment arm is due to interference bias. This would suggest that 54.16\% ($\pm65.05\%$) of the TATE achieved through a Bernoulli randomized experiment is due to interference that is eliminated by instead running a clustered experiment. However, none of these point estimates are statistically significant. A post-hoc power analysis of the algorithmic pricing experiment reveals that the meta-experiment is underpowered to detect reasonable effect sizes relative to the treatment effect estimated obtained in the Bernoulli randomized arm of the meta-experiment. Table \ref{tab:mde_meta} shows the calculated minimum detectable effect (MDE) for $\beta$, $\nu$, and $\xi$. Each of these MDEs is also overlaid in red on Figure \ref{fig:meta_analysis_plot_pricing_bookings_only}. Comparing the Bernoulli TATE estimate with the meta-experiment MDEs implies that interference bias would need to have approximately the same magnitude as our Bernoulli TATE estimate to be detectable.

This result highlights the difficulty of identifying (and reducing) interference bias using cluster randomization and meta-experimentation when the treatment intervention of interest is a behavioral nudge or some other type of intervention that will require ITT analysis. Although both the fee meta-experiment and the pricing meta-experiment have experimental samples of almost exactly the same size, one is able to detect statistically significant interference bias, while the other is not. Given that standard errors decrease with square root of the sample size, we estimate that a sample approximately 3.45 times as large would be required to detect interference bias in the algorithmic pricing meta-experiment.

\section{Discussion} \label{sec:discussion}

In this paper, we have taken the first empirical steps to understand the extent to which statistical inference can bias total average treatment effect estimates in online marketplace experiments. We have achieved this by presenting the results from two different pricing-related meta-experiments conducted on Airbnb, an online marketplace for accommodations and experiences. In each meta-experiment, some clusters of listings were assigned their experimental treatment using Bernoulli randomization, whereas others were assigned to their experimental treatment using cluster randomization. The motivation for our focus on pricing-related interventions was twofold; understanding customer price elasticities is crucial for both platform intermediaries and sellers, and there are strong reasons to suspect that pricing-relating experiments violate the stable unit treatment value assumption.

Analysis of our first meta-experiment, in which guest platform fees for treatment listings were increased relative to their peers in the control, provided clear evidence for interference bias in online marketplace experiments, and the potential for cluster randomization to mitigate this bias. While analysis of the Bernoulli meta-treatment arm alone suggested that the TATE was a decrease of 0.207 bookings per listing, a joint analysis of both meta-treatment arms revealed that 32.60\% of the reported TATE in the Bernoulli meta-treatment arm was due to interference bias that cluster randomization was able to eliminate. This figure represents a lower bound on the magnitude of interference bias, as our clusters likely do an imperfect job of capturing Airbnb listings that interfere with one another. While many recent papers measure the impact of innovative market mechanisms through field experiments \citep{horton.2015, filippas.2019}, very few of them explicitly account for interference bias. Based on our results, we argue that taking steps to reduce interference bias is crucial if researchers hope to estimate total average treatment effects accurately.

Analysis of the fee meta-experiment also reveals that the amount of bias in TATE estimates may depend on the extent to which a market is supply- or demand-constrained. Although our evidence is weak and comes from non-statistically significant point estimates, TATE estimates appear to be overstated due to interference bias to a greater extent in Airbnb markets that are demand constrained than in Airbnb markets that are supply constrained. Better understanding the relationships between supply elasticity, demand elasticity, and interference bias is a promising direction for future work. We also analyze a second meta-experiment, in which the treatment changes Airbnb hosts' algorithmically suggested prices, to understand how well our method can be applied to a behavioral nudge that requires ITT analysis. While point estimates suggest that the TATE estimate from the Bernoulli randomized meta-treatment arm is severely inflated due to interference bias, our results are not statistically significant, despite both meta-experiments having approximately equal sample sizes. This result highlights the difficulty of detecting interference bias for behavioral nudges and other treatment interventions that require ITT analysis. Unfortunately, these types of interventions are very common in online marketplaces. Future work might focus on developing even more sensitive tests for interference bias that will work more effectively when conducting such experiments.

In addition to cluster randomization, there are a number of analysis techniques that have been developed in the network experimentation literature, such as exposure modeling \citep{aronow.2012}, regression adjustment \citep{chin.2018}, and exact tests for interference \citep{athey.2018} that, if adopted to a commerce-based setting, could help to more accurately identify and reduce interference bias in online marketplace experiments. Furthermore, there are a number of open questions regarding how to best identify the sellers most likely to interfere with one another in an online marketplace setting. The clustering method described in this paper is by no means the only (or best) way to cluster sellers before performing cluster randomization. Higher quality clusters could lead to even greater interference bias reductions. Finally, while the approach described in this work can reduce bias due to interference between sellers, it does not consider the reduction of bias due to interference between \textit{buyers}. Given that, in general, online marketplaces have much less information about buyers, many of the approaches discussed thus far are unlikely to be effective. Developing methods that reduce interference bias on the buyer side of online marketplaces is a promising direction for future research.

Accounting for interference bias increases the logistical complexity of online marketplace experimentation. However, for many interventions, e.g., those that are designed to help platform intermediaries measure price elasticities, determining only the direction of a treatment effect is not sufficient; an accurate point estimate is required. Using pricing related meta-experiments on Airbnb as a test case, we have shown that interference bias can account for at least 32.60\% of a TATE estimate in an online marketplace experiment. In light of this result, we believe that accounting for interference bias can be worth the additional effort for many marketplace designers and researchers.

\clearpage
\section{Figures}

\begin{figure}[htpb]
  \centering
    \includegraphics[width=\textwidth]{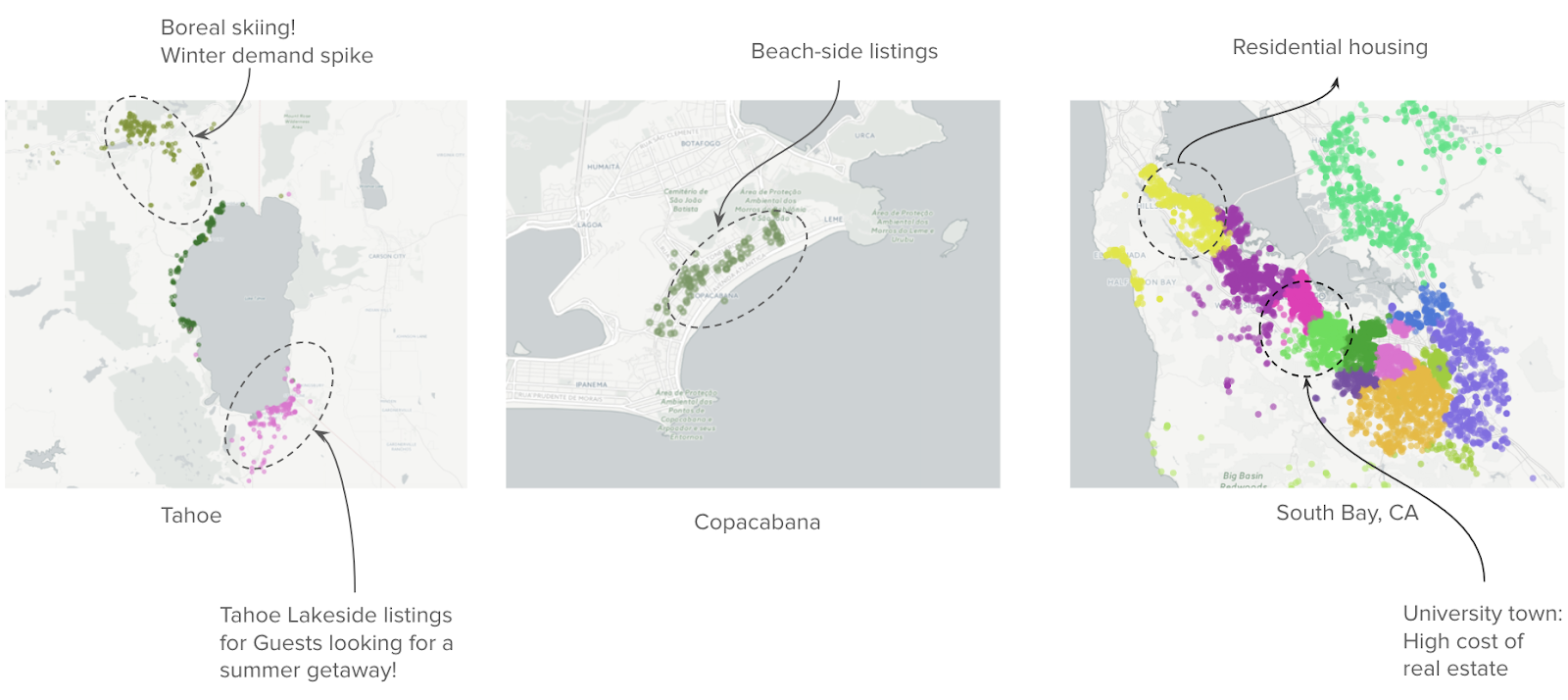}
      \caption{Example clusters generated using the hierarchical clustering scheme described in this paper. Image from \cite{srinivasan.2018}.}
      \label{fig:cluster_examples}
\end{figure}

\begin{figure}[htpb]
  \centering
    \includegraphics[width=\textwidth]{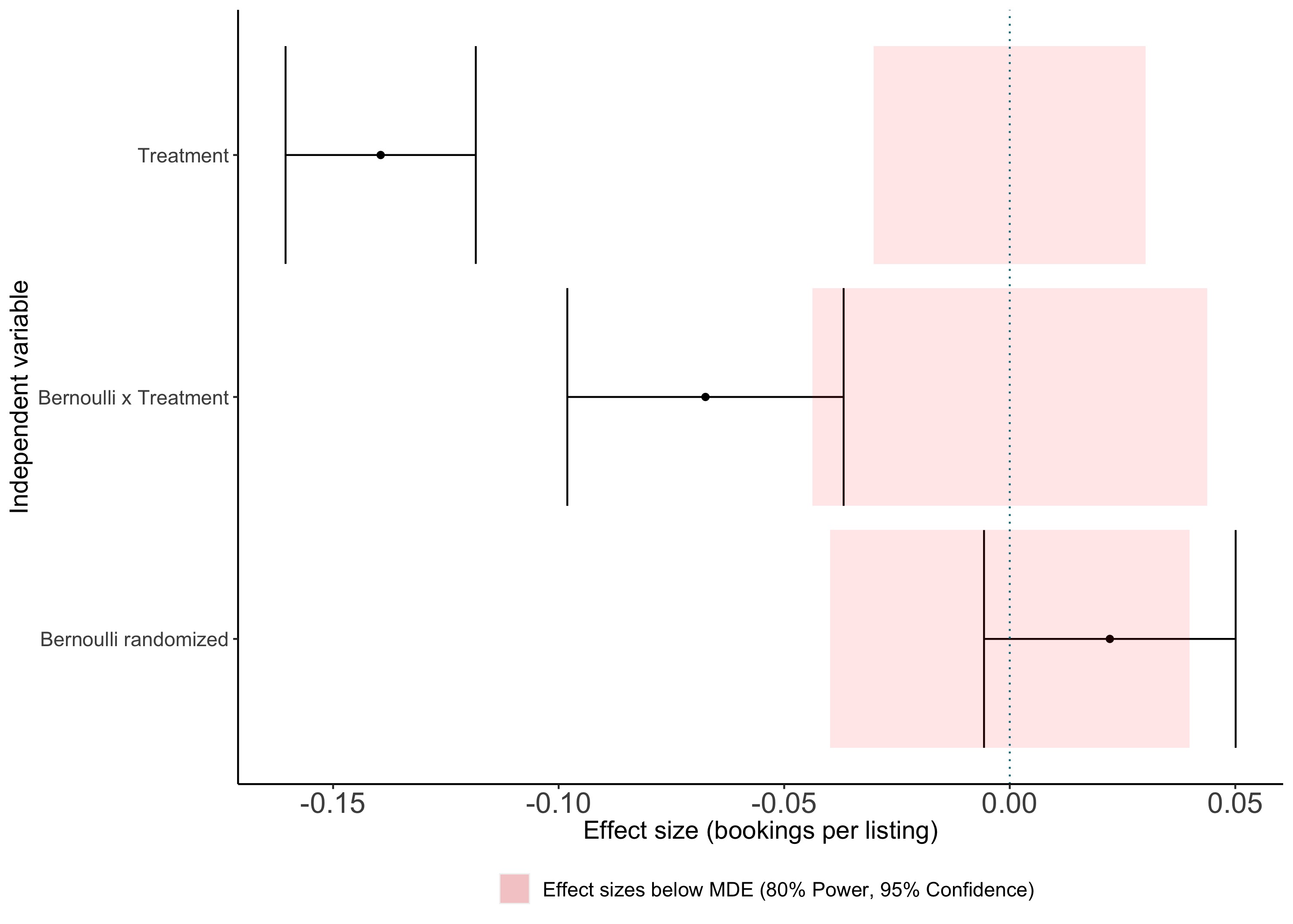}
      \caption{Coefficient estimates for the joint analysis of the fee meta-experiment. Error bars represent 95\% confidence intervals. The dotted blue line correponds to a treatment effect of 0 bookings per listing. The red shaded area corresponds to values that are below the MDE (80\% power, 95\% confidence).}
      \label{fig:meta_analysis_plot_fees_bookings_only}
\end{figure}

\clearpage
\section{Tables}

\begin{table}[!htbp] \centering 
  \caption{Independent results of the fee meta-experiment} 
  \label{tab:experiment_analysis_fees_bookings} 
\begin{tabular}{@{\extracolsep{5pt}}lcc} 
\\[-1.8ex]\hline 
\hline \\[-1.8ex] 
 & \multicolumn{2}{c}{\textit{Dependent variable:}} \\ 
\cline{2-3} 
\\[-1.8ex] & \multicolumn{2}{c}{Bookings} \\ 
 & Bernoulli randomized & Cluster randomized \\ 
\\[-1.8ex] & (1) & (2)\\ 
\hline \\[-1.8ex] 
 Treatment & $-$0.207$^{***}$ & $-$0.142$^{***}$ \\ 
  & (0.011) & (0.011) \\ 
  & & \\ 
 Pre-treatment bookings & 0.173$^{***}$ & 0.174$^{***}$ \\ 
  & (0.001) & (0.001) \\ 
  & & \\ 
 Pre-treatment nights booked & $-$0.003$^{***}$ & $-$0.003$^{***}$ \\ 
  & (0.000) & (0.000) \\ 
  & & \\ 
 Pre-treatment booking value & 0.000 & 0.000$^{***}$ \\ 
  & (0.000) & (0.000) \\ 
  & & \\ 
 Pre-treatment gross guest spend & $-$0.000$^{**}$ & $-$0.000$^{***}$ \\ 
  & (0.000) & (0.000) \\ 
  & & \\ 
\hline \\[-1.8ex] 
Stratum F.E. & Yes & Yes \\ 
Robust s.e. & Yes & Yes \\ 
Clustered s.e. & No & Yes \\ 
R$^{2}$ & 0.407 & 0.405 \\ 
Adjusted R$^{2}$ & 0.406 & 0.405 \\ 
\hline 
\hline \\[-1.8ex] 
\textit{Note:}  & \multicolumn{2}{r}{$^{*}$p$<$0.1; $^{**}$p$<$0.05; $^{***}$p$<$0.01} \\ 
\end{tabular} 
\end{table}

\begin{table}[!htbp] \centering 
  \caption{Results of the fees Meta-experiment} 
  \label{tab:meta_experiment_analysis_fees_bookings_only} 
\begin{tabular}{@{\extracolsep{5pt}}lc} 
\\[-1.8ex]\hline 
\hline \\[-1.8ex] 
 & \multicolumn{1}{c}{\textit{Dependent variable:}} \\ 
\cline{2-2} 
\\[-1.8ex] & Bookings \\ 
\hline \\[-1.8ex] 
 Treatment & $-$0.139$^{***}$ \\ 
  & (0.011) \\ 
  & \\ 
 Bernoulli Randomized & 0.022 \\ 
  & (0.014) \\ 
  & \\ 
 Bernoulli $\times$ Treatment & $-$0.067$^{***}$ \\ 
  & (0.016) \\ 
  & \\ 
 Pre-treatment bookings & 0.174$^{***}$ \\ 
  & (0.001) \\ 
  & \\ 
 Pre-treatment nights booked & $-$0.003$^{***}$ \\ 
  & (0.000) \\ 
  & \\ 
 Pre-treatment booking value & 0.000$^{***}$ \\ 
  & (0.000) \\ 
  & \\ 
 Pre-treatment gross guest spend & $-$0.000$^{***}$ \\ 
  & (0.000) \\ 
  & \\ 
\hline \\[-1.8ex] 
Stratum F.E. & Yes \\ 
Robust s.e. & Yes \\ 
Clustered s.e. & Yes \\ 
R$^{2}$ & 0.405 \\ 
Adjusted R$^{2}$ & 0.405 \\ 
\hline 
\hline \\[-1.8ex] 
\textit{Note:}  & \multicolumn{1}{r}{$^{*}$p$<$0.1; $^{**}$p$<$0.05; $^{***}$p$<$0.01} \\ 
\end{tabular} 
\end{table}

\begin{table}[!htbp] \centering 
  \caption{Results of the fee meta-experiment for supply- and demand-constrained listings} 
  \label{tab:meta_experiment_analysis_fees_supply_demand_constrained} 
\begin{tabular}{@{\extracolsep{5pt}}lcc} 
\\[-1.8ex]\hline 
\hline \\[-1.8ex] 
 & \multicolumn{2}{c}{\textit{Dependent variable:}} \\ 
\cline{2-3} 
\\[-1.8ex] & \multicolumn{2}{c}{Bookings} \\ 
 & Supply constrained & Demand constrained \\ 
\\[-1.8ex] & (1) & (2)\\ 
\hline \\[-1.8ex] 
 Treatment & $-$0.241$^{***}$ & $-$0.200$^{***}$ \\ 
  & (0.051) & (0.038) \\ 
  & & \\ 
 Bernoulli Randomized & $-$0.029 & $-$0.031 \\ 
  & (0.060) & (0.059) \\ 
  & & \\ 
 Bernoulli $\times$ Treatment & 0.052 & $-$0.036 \\ 
  & (0.059) & (0.052) \\ 
  & & \\ 
 Pre-treatment bookings & 0.170$^{***}$ & 0.174$^{***}$ \\ 
  & (0.002) & (0.002) \\ 
  & & \\ 
 Pre-treatment nights booked & $-$0.003$^{***}$ & $-$0.003$^{***}$ \\ 
  & (0.000) & (0.000) \\ 
  & & \\ 
 Pre-treatment booking value & 0.000 & 0.000$^{***}$ \\ 
  & (0.000) & (0.000) \\ 
  & & \\ 
 Pre-treatment gross guest spend & $-$0.000$^{**}$ & $-$0.000$^{***}$ \\ 
  & (0.000) & (0.000) \\ 
  & & \\ 
\hline \\[-1.8ex] 
Stratum F.E. & Yes & Yes \\ 
Robust s.e. & Yes & Yes \\ 
Clustered s.e. & Yes & Yes \\ 
R$^{2}$ & 0.421 & 0.389 \\ 
Adjusted R$^{2}$ & 0.420 & 0.388 \\ 
\hline 
\hline \\[-1.8ex] 
\textit{Note:}  & \multicolumn{2}{r}{$^{*}$p$<$0.1; $^{**}$p$<$0.05; $^{***}$p$<$0.01} \\ 
\end{tabular} 
\end{table}

\begin{table}[!htbp] \centering 
  \caption{Independent results of the algorithmic pricing meta-experiment} 
  \label{tab:experiment_analysis_pricing_bookings} 
\begin{tabular}{@{\extracolsep{5pt}}lcc} 
\\[-1.8ex]\hline 
\hline \\[-1.8ex] 
 & \multicolumn{2}{c}{\textit{Dependent variable:}} \\ 
\cline{2-3} 
\\[-1.8ex] & \multicolumn{2}{c}{Bookings} \\ 
 & Bernoulli randomized & Cluster randomized \\ 
\\[-1.8ex] & (1) & (2)\\ 
\hline \\[-1.8ex] 
 Treatment & $-$0.106$^{***}$ & $-$0.051$^{*}$ \\ 
  & (0.028) & (0.029) \\ 
  & & \\ 
 Pre-treatment bookings & 0.822$^{***}$ & 0.828$^{***}$ \\ 
  & (0.004) & (0.002) \\ 
  & & \\ 
 Pre-treatment nights booked & $-$0.018$^{***}$ & $-$0.017$^{***}$ \\ 
  & (0.001) & (0.000) \\ 
  & & \\ 
 Pre-treatment booking value & 0.000$^{*}$ & 0.000$^{***}$ \\ 
  & (0.000) & (0.000) \\ 
  & & \\ 
 Pre-treatment gross guest spend & $-$0.000$^{**}$ & $-$0.000$^{***}$ \\ 
  & (0.000) & (0.000) \\ 
  & & \\ 
 Smart pricing pre-treatment & 0.587$^{***}$ & 0.586$^{***}$ \\ 
  & (0.033) & (0.020) \\ 
  & & \\ 
\hline \\[-1.8ex] 
Stratum F.E. & Yes & Yes \\ 
Robust s.e. & Yes & Yes \\ 
Clustered s.e. & No & Yes \\ 
R$^{2}$ & 0.580 & 0.578 \\ 
Adjusted R$^{2}$ & 0.578 & 0.578 \\ 
\hline 
\hline \\[-1.8ex] 
\textit{Note:}  & \multicolumn{2}{r}{$^{*}$p$<$0.1; $^{**}$p$<$0.05; $^{***}$p$<$0.01} \\ 
\end{tabular} 
\end{table}

\begin{table}[!htbp] \centering 
  \caption{Results of the algorithmic pricing meta-experiment} 
  \label{tab:meta_experiment_analysis_pricing_bookings_only} 
\begin{tabular}{@{\extracolsep{5pt}}lc} 
\\[-1.8ex]\hline 
\hline \\[-1.8ex] 
 & \multicolumn{1}{c}{\textit{Dependent variable:}} \\ 
\cline{2-2} 
\\[-1.8ex] & Bookings \\ 
\hline \\[-1.8ex] 
 Treatment & $-$0.050$^{*}$ \\ 
  & (0.030) \\ 
  & \\ 
 Bernoulli Randomized & $-$0.013 \\ 
  & (0.037) \\ 
  & \\ 
 Bernoulli $\times$ Treatment & $-$0.059 \\ 
  & (0.041) \\ 
  & \\ 
 Pre-treatment bookings & 0.827$^{***}$ \\ 
  & (0.002) \\ 
  & \\ 
 Pre-treatment nights booked & $-$0.017$^{***}$ \\ 
  & (0.000) \\ 
  & \\ 
 Pre-treatment booking value & 0.000$^{***}$ \\ 
  & (0.000) \\ 
  & \\ 
 Pre-treatment gross guest spend & $-$0.000$^{***}$ \\ 
  & (0.000) \\ 
  & \\ 
 Smart pricing pre-treatment & 0.577$^{***}$ \\ 
  & (0.017) \\ 
  & \\ 
\hline \\[-1.8ex] 
Stratum F.E. & Yes \\ 
Robust s.e. & Yes \\ 
Clustered s.e. & Yes \\ 
R$^{2}$ & 0.577 \\ 
Adjusted R$^{2}$ & 0.577 \\ 
\hline 
\hline \\[-1.8ex] 
\textit{Note:}  & \multicolumn{1}{r}{$^{*}$p$<$0.1; $^{**}$p$<$0.05; $^{***}$p$<$0.01} \\ 
\end{tabular} 
\end{table} 

\begin{table}[ht]
\centering
\caption{Minimum detectable effects for algorithmic pricing meta-experiment analysis (power = 80\%, confidence level = 95\%)} 
\label{tab:mde_meta}
\begin{tabular}{lr}
  \hline
Regressor & Bookings \\ 
  \hline
Treatment & 0.084 \\ 
  Bernoulli x Treatment & 0.114 \\ 
  Bernoulli randomized & 0.082 \\ 
   \hline
\end{tabular}
\end{table}

\clearpage

%
%
%

\clearpage

\bibliographystyle{informs2014} 
\bibliography{pricing_meta_experiment} 

\clearpage

\renewcommand\thefigure{\thesection.\arabic{figure}}    
\setcounter{figure}{0}  

\renewcommand\thetable{\thesection.\arabic{table}}    
\setcounter{table}{0}  

\begin{APPENDICES}

\section{Method for cluster size selection} \label{sec:cluster_size_selection}

In this section, we detail the methodology that was used in deciding to conduct the fee meta-experiment with clusters with a listing threshold of 1,000, as opposed to 250. Although this analysis was originally conducted using clusters and data from February 2019, we present analyses using clusters generated on January 5, 2020, listing views occurring between January 5, 2020 and January 12, 2020, and bookings occurring between January 5, 2020 and January 26, 2020. However, the results we report and the corresponding conclusions are qualitatively similar to those obtained using 2019 data.

In choosing a cluster size threshold, the fundamental trade-off is between statistical power and capturing Airbnb demand. While smaller clusters will yield more statistical power (since there will be more of them), they will also do a poorer job of capturing demand, since a given user search session is more likely to contain listings from many different clusters. On the other hand, larger clusters will provide less statistical power, but will do a better job of capturing demand. Power analysis suggested that a week-long experiment shifting fees in the same manner as our fee experiment would have an MDE of 0.9\% for interference bias if clusters with a threshold size of 250 were used, whereas the same experiment would have an MDE of 1.05\% for interference bias if clusters with a threshold size of 1,000 were used. In order to determine whether this reduction in ``ideal" MDE is worthwhile, we needed to measure differences in the extent to which the two sets of clusters capture demand.


We began our investigation by defining two different measures related to demand capture:

\begin{equation}
\textrm{\% in single cluster } = \frac{1}{n_{users}} \sum_{\textrm{all users}} \mathbbm{1} \left ( n_{clusters} = 1 \right )
\end{equation}

\begin{equation}
\textrm{Demand capture } = \frac{1}{n_{users}} \sum_{\textrm{all users}} \left(1 - \frac{n_{clusters}}{n_{listings}} \right)
\end{equation}

\noindent The first measures the percentage of users for whom all listings viewed fall within a single cluster. The second is a less strict measure that captures the extent to which all viewed listings are contained within a small number of clusters. Importantly, both measures will be close or equal to 1 if users never compare listings across different clusters and $n_{listings}$ is sufficiently large, and will be equal to 0 if the number of listings a user compares is equal to the number of clusters needed to cover them. Figure \ref{fig:cluster_coverage_comparison_plot} shows both of these measures for listing views occurring between January 5, 2020 and January 12, 2020, for cluster size thresholds ranging from 100 to entire markets. As expected, as the size of clusters increases, both of these demand capture metrics move closer to 1. Importantly, even when markets are defined as ``clusters," they are unable to capture 100\% of demand, regardless of which measure we use.

Based on statistical power considerations, we decided that a cluster size threshold of 1,000 was the maximum threshold worth considering. Once this decision was made, we began to more directly compare the status quo threshold of 250 listings (which had been used for the algorithmic pricing meta-experiment) to the maximum threshold of 1,000 listings.\footnote{The 250 listing threshold was chosen for the algorithmic pricing meta-experiment in an ad-hoc manner.} In doing so, we created an alternative demand capture measure that asked the following question: given a set of clusters, what percentage of listing viewers have at least $x\%$ of their listings captured by one cluster? Figure \ref{fig:single_cluster_coverage_plot} plots this measure for both the 250 listing threshold clusters and the 1,000 listing threshold clusters, with demand capture thresholds of 67\%, 75\%, and 90\%. As expected, the clusters with the 1,000 listing threshold do a better job of capturing demand than the 250 listing threshold clusters.

In order to make a principled decision, we assumed that the ``ideal" MDEs mentioned earlier in this appendix were reduced by poor demand capture according to the relationship below:

\begin{equation}
MDE_{actual} = \frac{MDE_{ideal}}{\textrm{Demand capture}}.
\end{equation}

\noindent In other words, as a given set of clusters' demand capture moved closer to 1, the MDE would approach the ideal MDE. Given this assumed relationship between actual MDE, ideal MDE, and demand capture, we determined that the 1,000 listing threshold clusters would be preferable to the 250 listing threshold clusters when

\begin{equation}
\frac{\textrm{Demand capture}_{1,000}}{\textrm{Demand capture}_{250}} > \frac{MDE_{ideal_{250}}}{MDE_{ideal_{1,000}}} \rightarrow 
\frac{\textrm{Demand capture}_{1,000}}{\textrm{Demand capture}_{250}} > \frac{1.05\%}{0.9\%}
\end{equation}

Table \ref{tab:cluster_size_comparison} shows the ratio of demand capture for clusters with a threshold of 1,000 listings to the demand capture for clusters with a threshold of 250 clusters according to five different demand capture measures: the average share of search listings belonging to a cluster, the average user-level Herfindahl-Hirschman index across clusters, and the percentage of users for which one cluster accounts for at least 67\%, 75\%, and 90\% of listings viewed. Across all five of these demand capture metrics, and across different user subpopulations, the demand capture ratio is consistently above $\frac{1.05\%}{0.9\%} = 1.17$. Based on this calculation, we determined that clusters with a threshold of 1,000 listings were preferable.

\section{Interference bias for nights booked and gross guest spend} \label{sec:nb_and_bv}

In addition to bookings per listing, we also conducted the main analyses in our paper for both nights booked per listing and gross guest spend per listing. In this appendix, we present the results of our analyses for these additional outcomes. Qualitatively, our results for nights booked per listing and gross guest spend per listing are extremely similar to our results for bookings per listing.

\subsection{Fee meta-experiment}

Table \ref{tab:experiment_analysis_fees_no_bookings} shows the estimated effect of the fee treatment in both the Bernoulli randomized meta-treatment arm and the cluster randomized meta-treatment arm on both nights booked per listing and gross guest spend per listing. Our TATE estimates for each outcome are also depicted, along with 95\% confidence intervals, in Figure \ref{fig:arm_level_results_fees_no_bookings}. We estimate in the Bernoulli randomized meta-treatment arm that the treatment led to a statistically significant loss of 0.768 nights booked per listing and \$79.68 in gross guest spend per listing, whereas we estimate in the cluster randomized meta-treatment arm that the treatment led to a statistically significant loss of 0.579 nights booked per listing and \$63.39 in booking value per listing.

In order to test whether or not there is a statistically significant difference between the TATE estimates in the two meta-treatment arms, we conduct a joint analysis of both meta-treatment arms simultaneously. Table \ref{tab:meta_experiment_analysis_fees_no_bookings} shows our results. Our results are also depicted in Figure \ref{fig:meta_analysis_plot_fees_no_bookings.jpg}, along with 95\% confidence intervals. We find statistically significant evidence of interference bias in the Bernoulli TATE estimate for nights booked per listing at the 95\% confidence level, but do not find statistically significant evidence of interference bias in the Bernoulli TATE estimate for gross guest spend per listing. Our point estimates suggest that interference accounts for 24.79\% of the Bernoulli TATE estimate for nights booked per listing (stat sig.) and 21.04\% of the Bernoulli TATE estimate for gross guest spend per listing (not stat. sig).

\subsection{Algorithmic pricing meta-experiment}

Table \ref{tab:experiment_analysis_pricing_no_bookings} shows the estimated effect of the algorithmic pricing treatment in both the Bernoulli randomized meta-treatment arm and the cluster randomized meta-treatment arm on both nights booked per listing and gross guest spend per listing. Our TATE estimates for each outcome are depicted, along with 95\% confidence intervals, in Figure \ref{fig:arm_level_results_pricing_no_bookings},  We estimate in the Bernoulli randomized meta-treatment arm that the treatment let do a statistically significant loss of 0.288 nights booked per listing and \$37.38 in gross guest spend per listing, whereas we do not detect a statistically significant treatment effect for either outcome in the cluster randomized meta-treatment arm.

In order to test whether or not there is a statistically significant difference between the TATE estimates in the two meta-treatment arms, we conduct a joint analysis of both meta-treatment arms simultaneously. Table \ref{tab:meta_experiment_analysis_pricing_no_bookings} shows our results. Our results are also depicted in Figure \ref{fig:meta_analysis_plot_pricing_no_bookings.jpg}, along with 95\% confidence intervals. We do not find statistically significant evidence for interference bias for either outcome. While not statistically significant, our point estimates suggest that interference accounts for 36.86\% of the Bernoulli TATE estimate for nights booked per listing and 104.73\% of the Bernoulli TATE estimate for gross guest spend per listing.

\section{Cluster-level analysis of cluster-randomized meta-treatment arm} \label{sec:cluster_agg_analysis}

Rather than analyzing data from the cluster randomized meta-treatment arm of our experiments at the individual level with clustered standard errors, it is also possible to aggregate data at the \textit{cluster} level and instead estimated a weighted version of Equation \ref{eq:main_experiment_spec}, where each cluster is weighted according to the number of experiment-eligible listings in that cluster. In this appendix, we compare the cluster randomized TATE estimates obtained using these two different approaches.

\subsection{Fee meta-experiment}

Table \ref{tab:cluster_experiment_analysis_fees_compare} compares the TATE estimates obtained from the cluster randomized meta-treatment arm of the fee meta-experiment when analyzing the data at both the individual listing level and at the cluster level. Our results are also depicted in Figure \ref{fig:arm_level_results_fees_no_bookings_compare_clus}. We find that both approaches yield almost identical TATE point estimates and standard errors.

\subsection{Algorithmic pricing meta-experiment}

Table \ref{tab:cluster_experiment_analysis_pricing_compare} compares the TATE estimates obtained from the cluster randomized meta-treatment arm of the algorithmic pricing meta-experiment when analyzing the data at both the individual listing level and at the cluster level. Our results are also depicted in Figure \ref{fig:arm_level_results_pricing_no_bookings_compare_clus}. We find that both approaches yield almost identical TATE point estimates and standard errors.

\section{Results with mixed units of analysis} \label{sec:mixed_analysis}

In addition to performing joint analysis of our meta-experiments with listing-level data, it is possible to analyze the meta-experiments with a mixture of listing-level data and data aggregated at the cluster level. For both meta-experiments, we estimate Equation \ref{eq:meta_experiment_spec} on listing-level data from the Bernoulli randomized meta-treatment arm and cluster-level data from the cluster randomized meta-treatment arm. Observations are weighted by the number of listings making up that observation (i.e., listings receive a weight of 1, whereas clusters receive a weight equal to the number of experiment eligible listings in that cluster). In this appendix, we compare results obtained using this approach with those obtained analyzing the meta-experiment entirely with listing level data.

\subsection{Fee meta-experiment}

Table \ref{tab:meta_experiment_analysis_fees_bookings_only_agg_comp} compares results obtained by analyzing the fee meta-experiment at the listing level and with mixed units of analysis. Our results are also depicted in Figure \ref{fig:meta_analysis_plot_bookings_only_comp_agg}. We find that both approaches yield almost identical results.

\subsection{Algorithmic pricing meta-experiment}

Table \ref{tab:meta_experiment_analysis_pricing_bookings_only_agg_comp} compares results obtained by analyzing the fee meta-experiment at the listing level and with mixed units of analysis. Our results are also depicted in Figure \ref{fig:meta_analysis_plot_bookings_only_comp_agg_pricing}. We find that both approaches yield almost identical results.

\clearpage
\section{Additional Figures}

\begin{figure}[htpb]
  \centering
    \includegraphics[width=\textwidth]{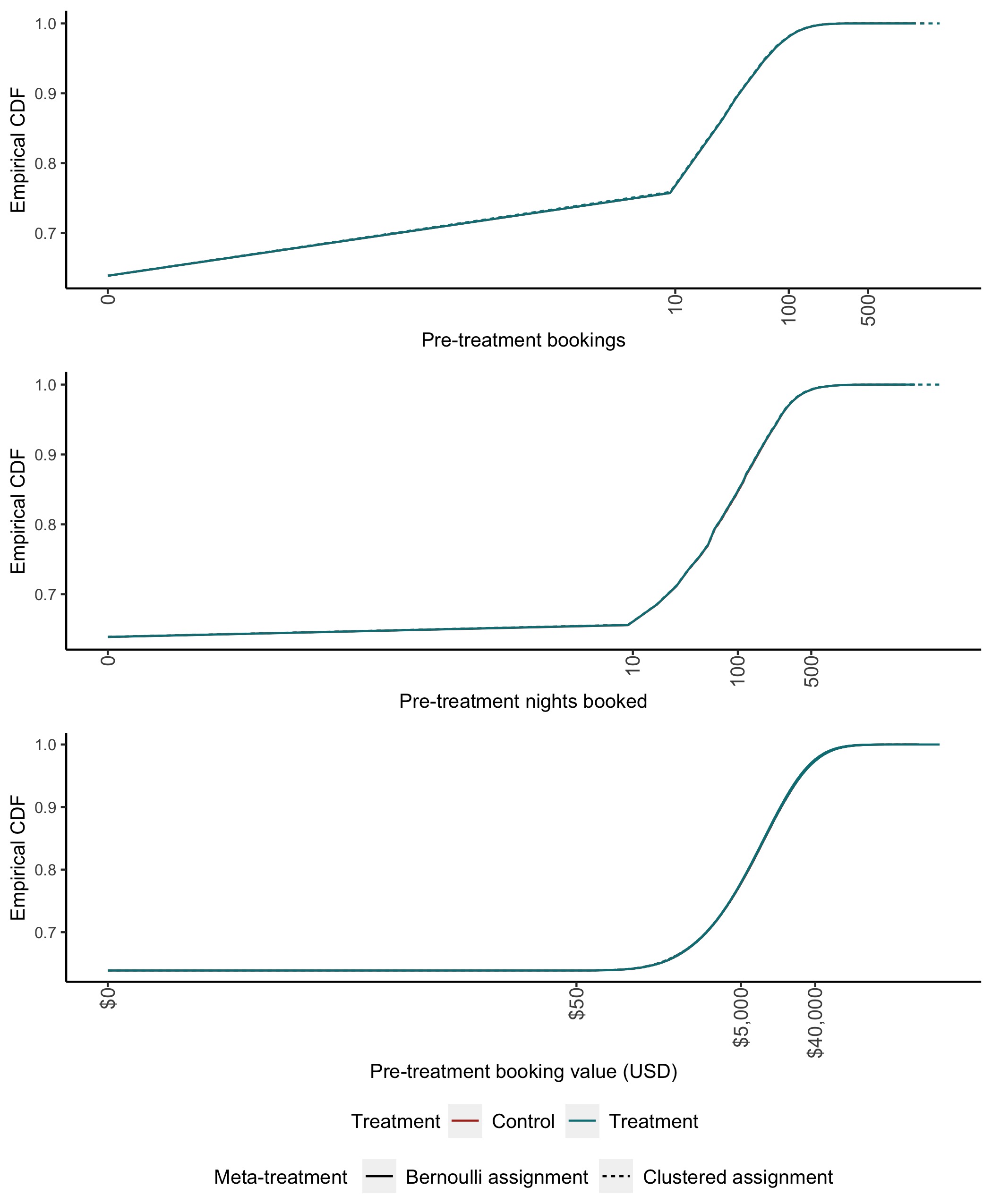}
      \caption{The empirical CDFs for pre-treatment bookings, nights booked, and booking value in each of the four treatment/meta-treatment groups for the fee meta-experiment.}
      \label{fig:ecdf_fee_experiment}
\end{figure}

\begin{figure}[htpb]
  \centering
    \includegraphics[width=\textwidth]{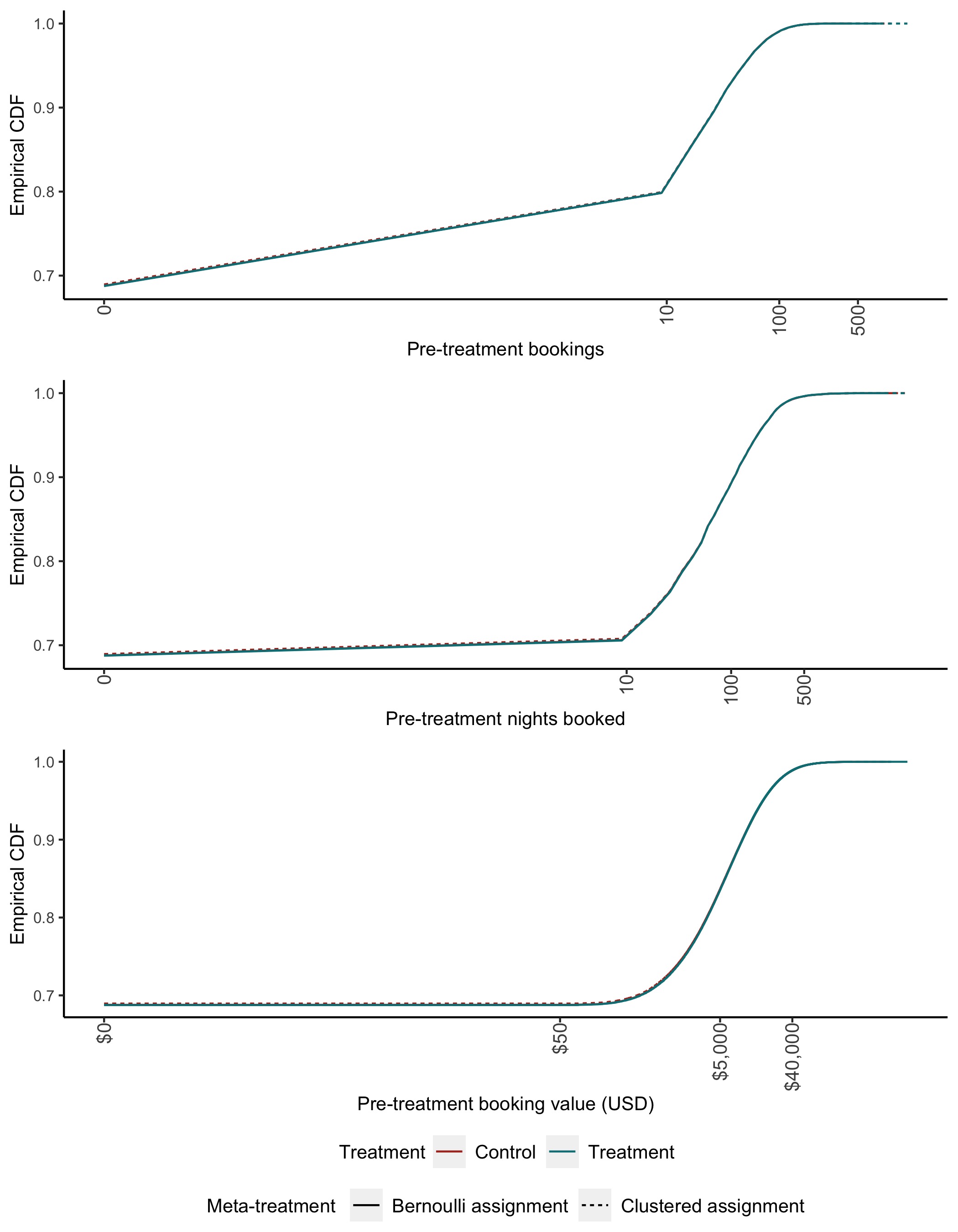}
      \caption{The empirical CDFs for pre-treatment bookings, nights booked, and booking value in each of the four treatment/meta-treatment groups for the algorithmic pricing meta-experiment.}
      \label{fig:ecdf_pricing_experiment}
\end{figure}

\begin{figure}[htpb]
 \centering
    \includegraphics[width=\textwidth]{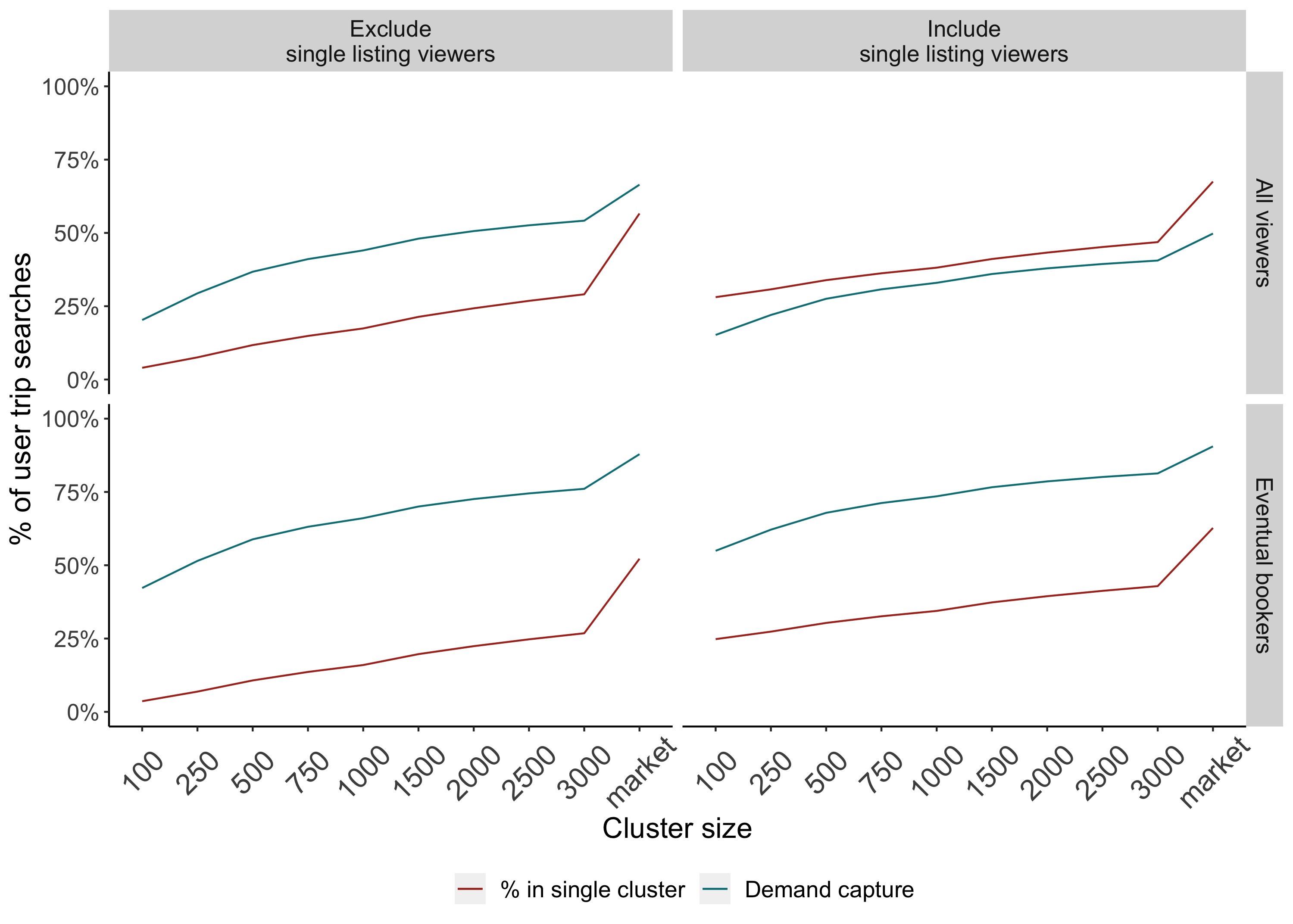}
      \caption{The relationship between cluster size and demand capture for two different metrics. The left column excludes users who only view a single Airbnb listing, whereas the right column includes them. The top row includes all listing viewers, whereas the bottom row only includes Airbnb users who go on to eventually book a listing.}
      \label{fig:cluster_coverage_comparison_plot}
\end{figure}

\begin{figure}[htpb]
 \centering
    \includegraphics[width=\textwidth]{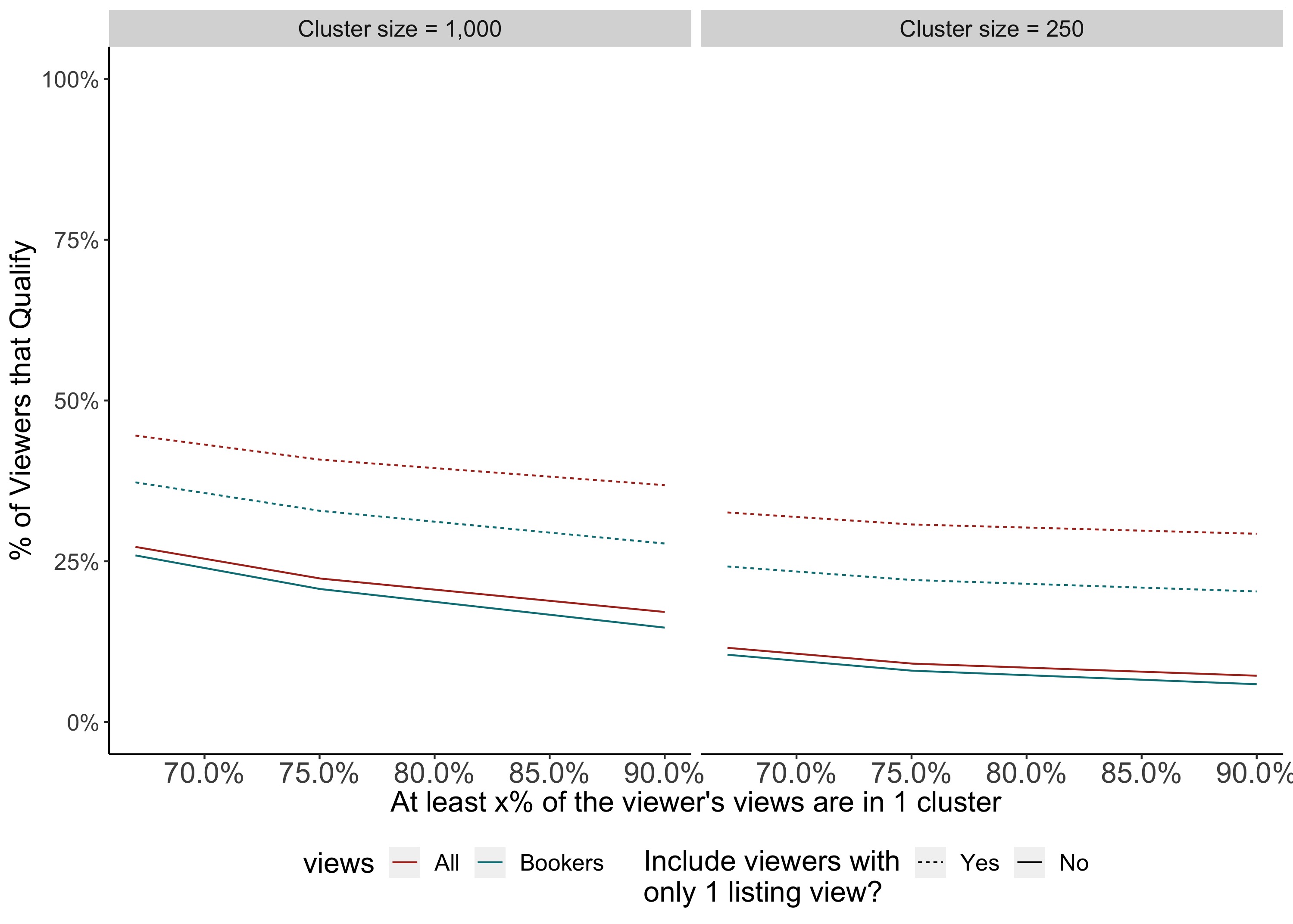}
      \caption{A direct comparison of the demand capture of clusters with a 1,000 listing threshold, and clusters with a 250 listing threshold. Curves show the percentage of viewers for whom at least $x\%$ of their views are contained by one cluster. Red curves include all listing viewers, whereas blue curves only include Airbnb users who go on to eventually book a listing. Dashed lines include users who only view a single Airbnb listing, whereas solid lines do not.}
      \label{fig:single_cluster_coverage_plot}
\end{figure}

\begin{figure}[htpb]
 \centering
    \includegraphics[width=\textwidth]{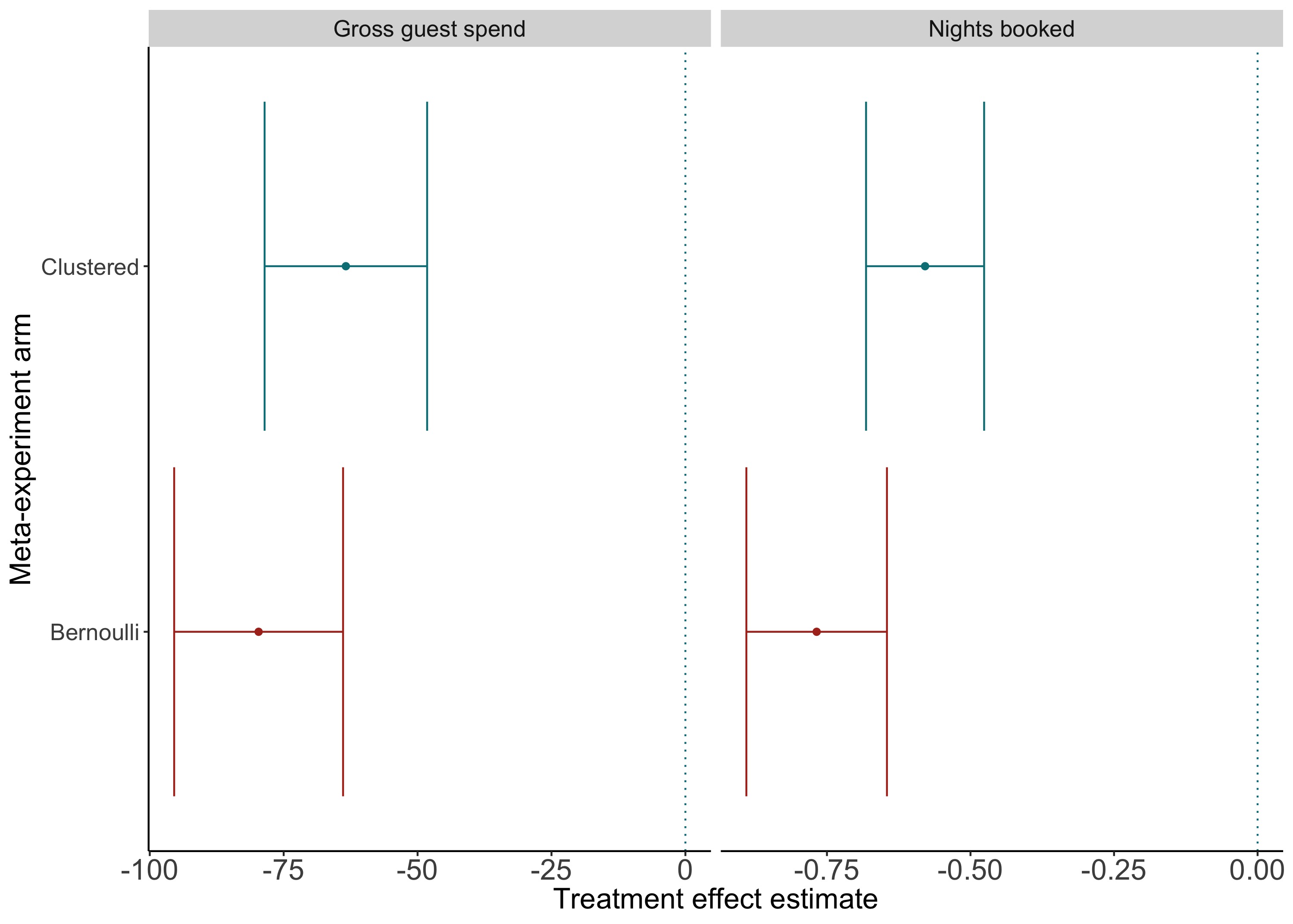}
      \caption{Total average treatment effect estimates (nights booked per listing and gross guest spend per listing) for the fee experiment, estimated separately in the Bernoulli randomized meta-treatment arm and the cluster randomized meta treatment arm. Error bars represent 95\% confidence intervals. The dotted blue line corresponds to a treatment effect of 0.}
      \label{fig:arm_level_results_fees_no_bookings}
\end{figure}

\begin{figure}[htpb]
 \centering
    \includegraphics[width=\textwidth]{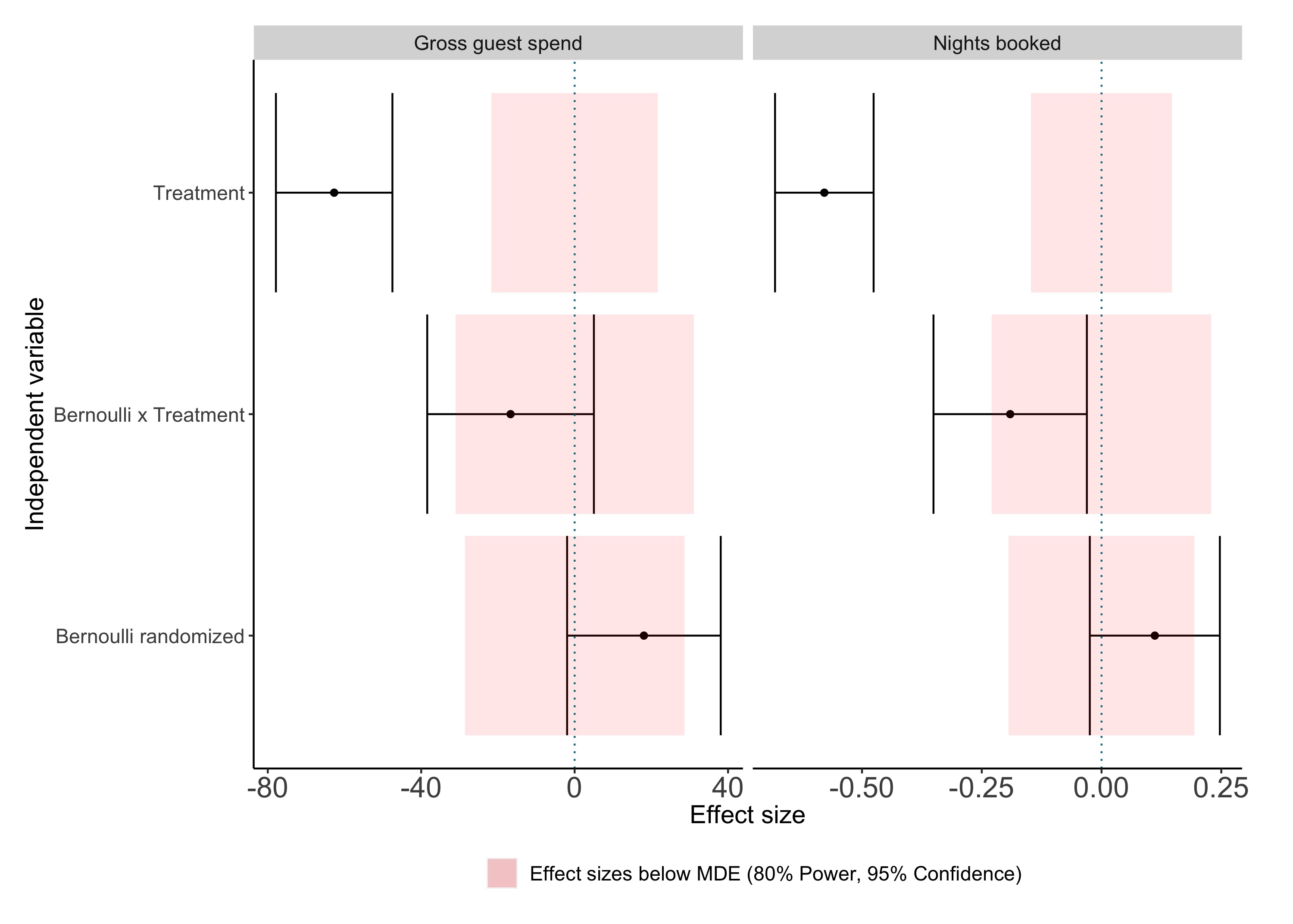}
      \caption{Coefficient estimates for the joint analysis of the fee meta-experiment (nights booked per listing and gross guest spend per listing). Error bars represent 95\% confidence intervals. The dotted blue line corresponds to a treatment effect of 0. The red shaded area corresponds to values that are below the MDE (80\% power, 95\% confidence).}
      \label{fig:meta_analysis_plot_fees_no_bookings.jpg}
\end{figure}

\begin{figure}[htpb]
 \centering
    \includegraphics[width=\textwidth]{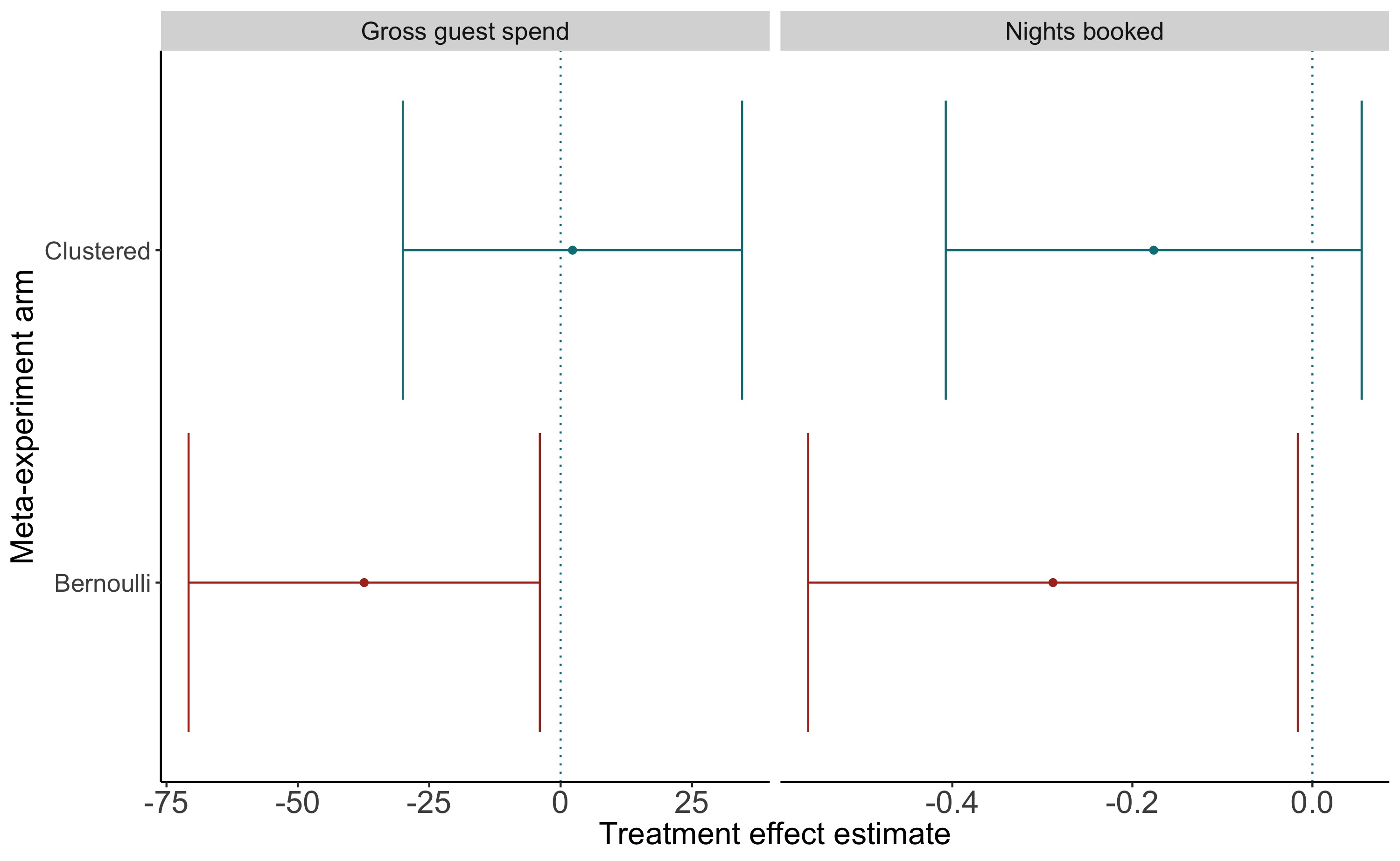}
      \caption{Total average treatment effect estimates (nights booked per listing and gross guest spend per listing) for the algorithmic pricing experiment, estimated separately in the Bernoulli randomized meta-treatment arm and the cluster randomized meta treatment arm. Error bars represent 95\% confidence intervals. The dotted blue line corresponds to a treatment effect of 0.}
      \label{fig:arm_level_results_pricing_no_bookings}
\end{figure}

\begin{figure}[htpb]
 \centering
    \includegraphics[width=\textwidth]{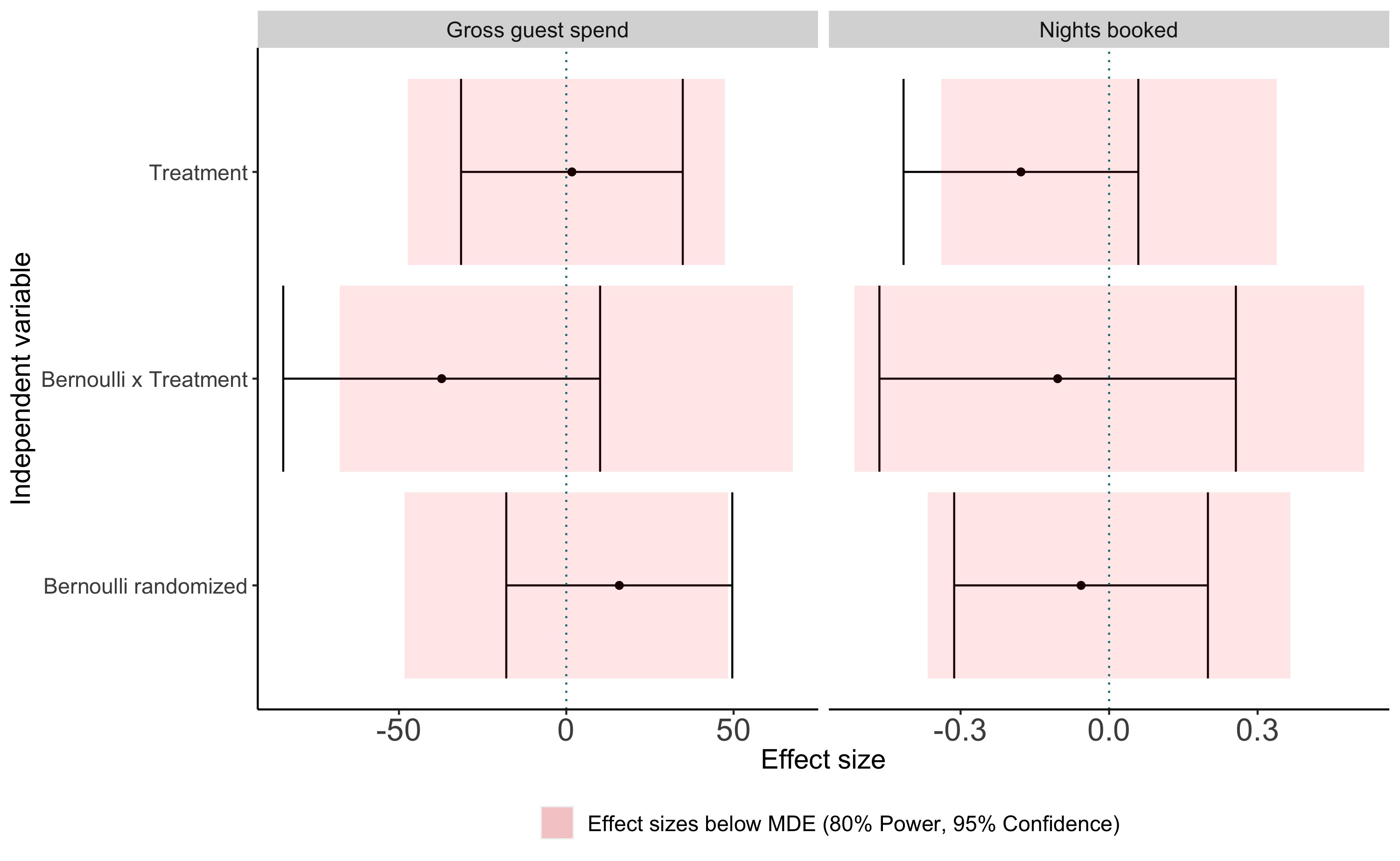}
      \caption{Coefficient estimates for the joint analysis of the algorithmic pricing meta-experiment (nights booked per listing and gross guest spend per listing). Error bars represent 95\% confidence intervals. The dotted blue line corresponds to a treatment effect of 0. The red shaded area corresponds to values that are below the MDE (80\% power, 95\% confidence).}
      \label{fig:meta_analysis_plot_pricing_no_bookings.jpg}
\end{figure}

\begin{figure}[htpb]
 \centering
    \includegraphics[width=\textwidth]{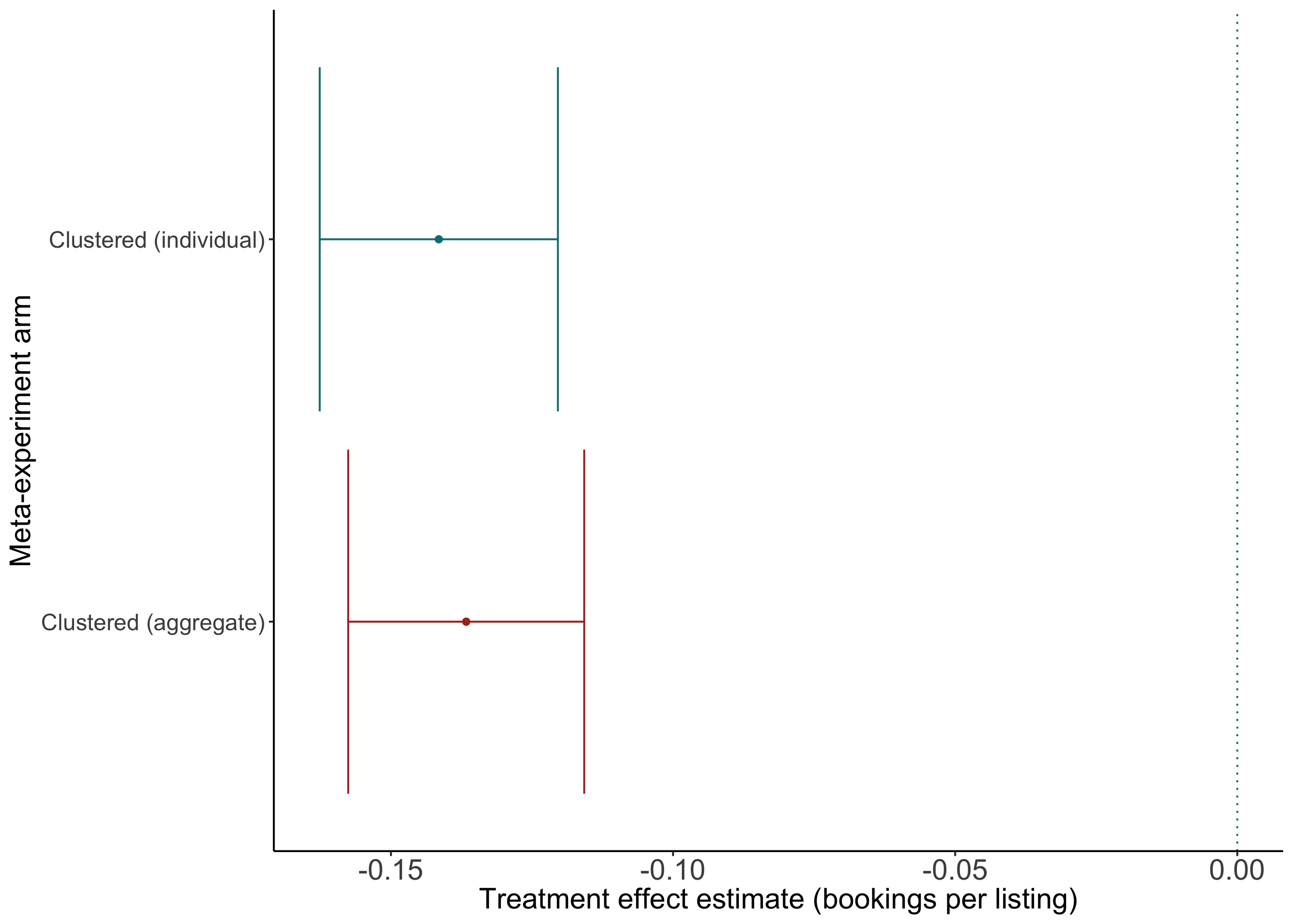}
      \caption{Comparison of the TATE estimates from the cluster randomized meta-treatment arm of the fees experiment, obtained analyzing data at either the individual listing level or at the cluster level. Error bars represent 95\% confidence intervals. The dotted blue line corresponds to a treatment effect of 0 bookings per listing.}
      \label{fig:arm_level_results_fees_no_bookings_compare_clus}
\end{figure}

\begin{figure}[htpb]
 \centering
    \includegraphics[width=\textwidth]{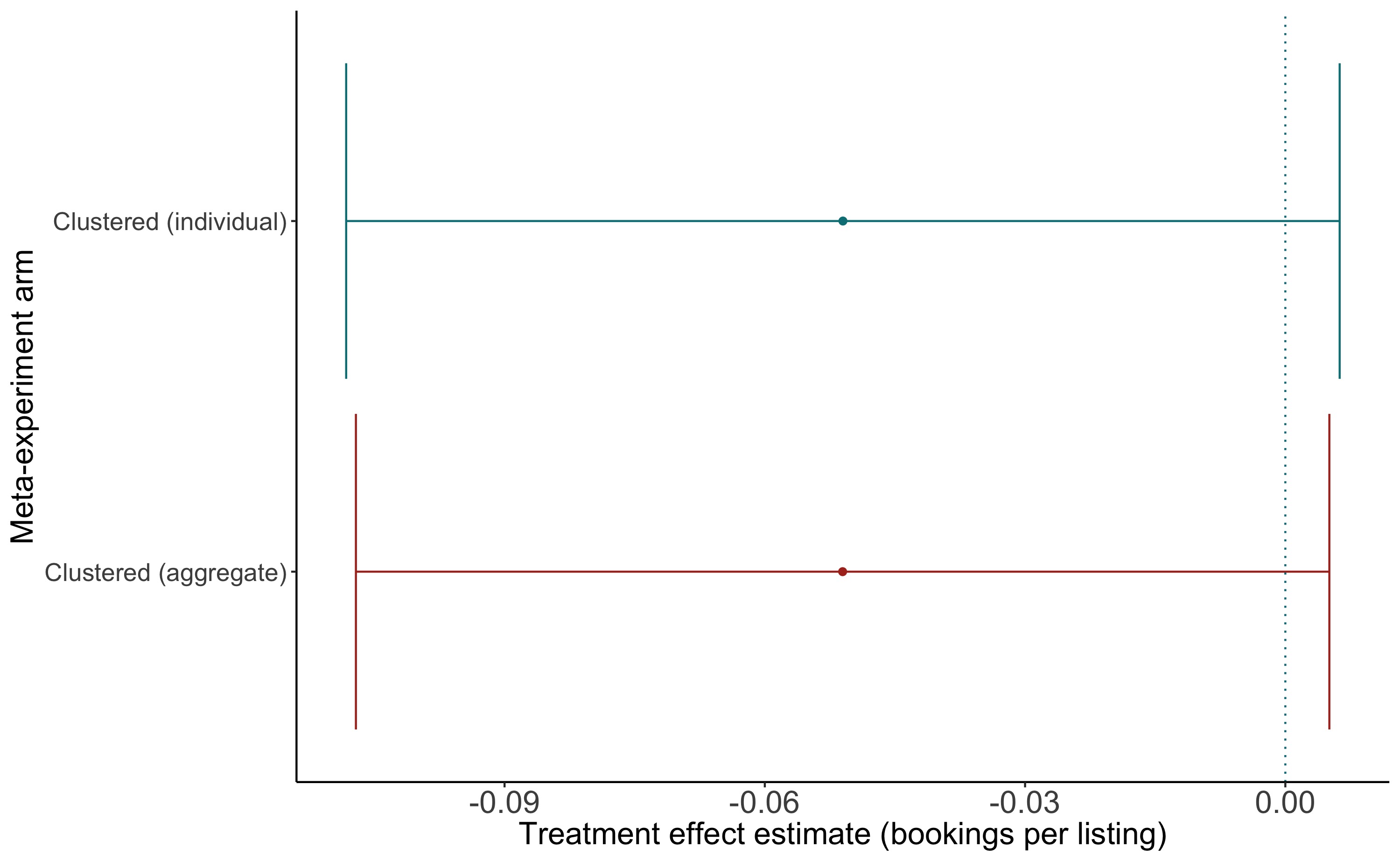}
      \caption{Comparison of the TATE estimates from the cluster randomized meta-treatment arm of the algorithmic pricing experiment, obtained analyzing data at either the individual listing level or at the cluster level. Error bars represent 95\% confidence intervals. The dotted blue line corresponds to a treatment effect of 0 bookings per listing.}
      \label{fig:arm_level_results_pricing_no_bookings_compare_clus}
\end{figure}

\begin{figure}[htpb]
 \centering
    \includegraphics[width=\textwidth]{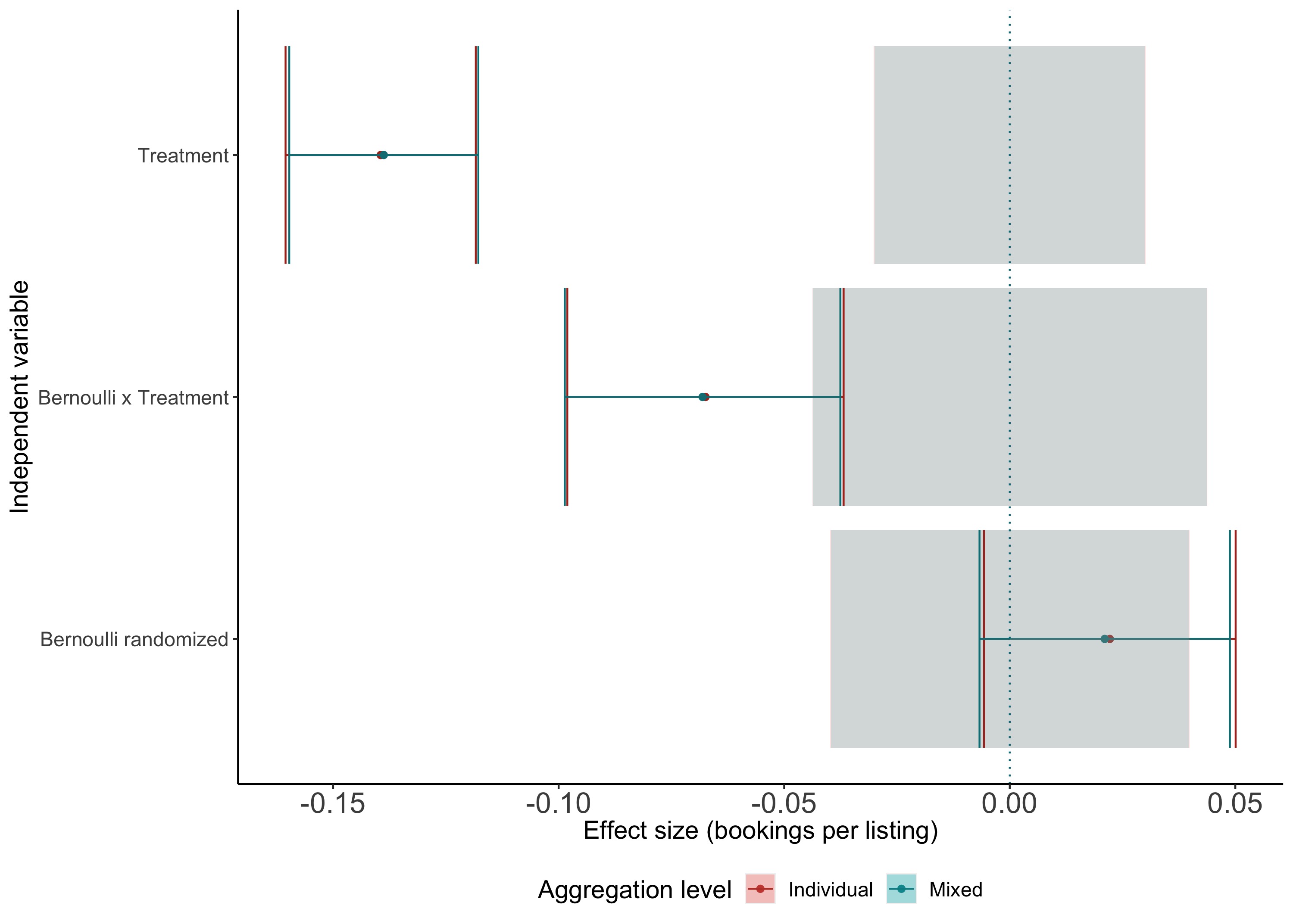}
      \caption{Comparison of fee experiment meta-analysis estimates obtained analyzing data at the individual level of analysis, and the mixed level of analysis. In the mixed analysis, Data from Bernoulli randomized listings is included at the listing level, whereas data from cluster randomized listings is aggregated at the cluster level. Error bars correspond to 95\% confidence intervals. Shaded areas represent effect sizes below the MDE threshold (80\% power, 95\% confidence).}
      \label{fig:meta_analysis_plot_bookings_only_comp_agg}
\end{figure}

\begin{figure}[htpb]
 \centering
    \includegraphics[width=\textwidth]{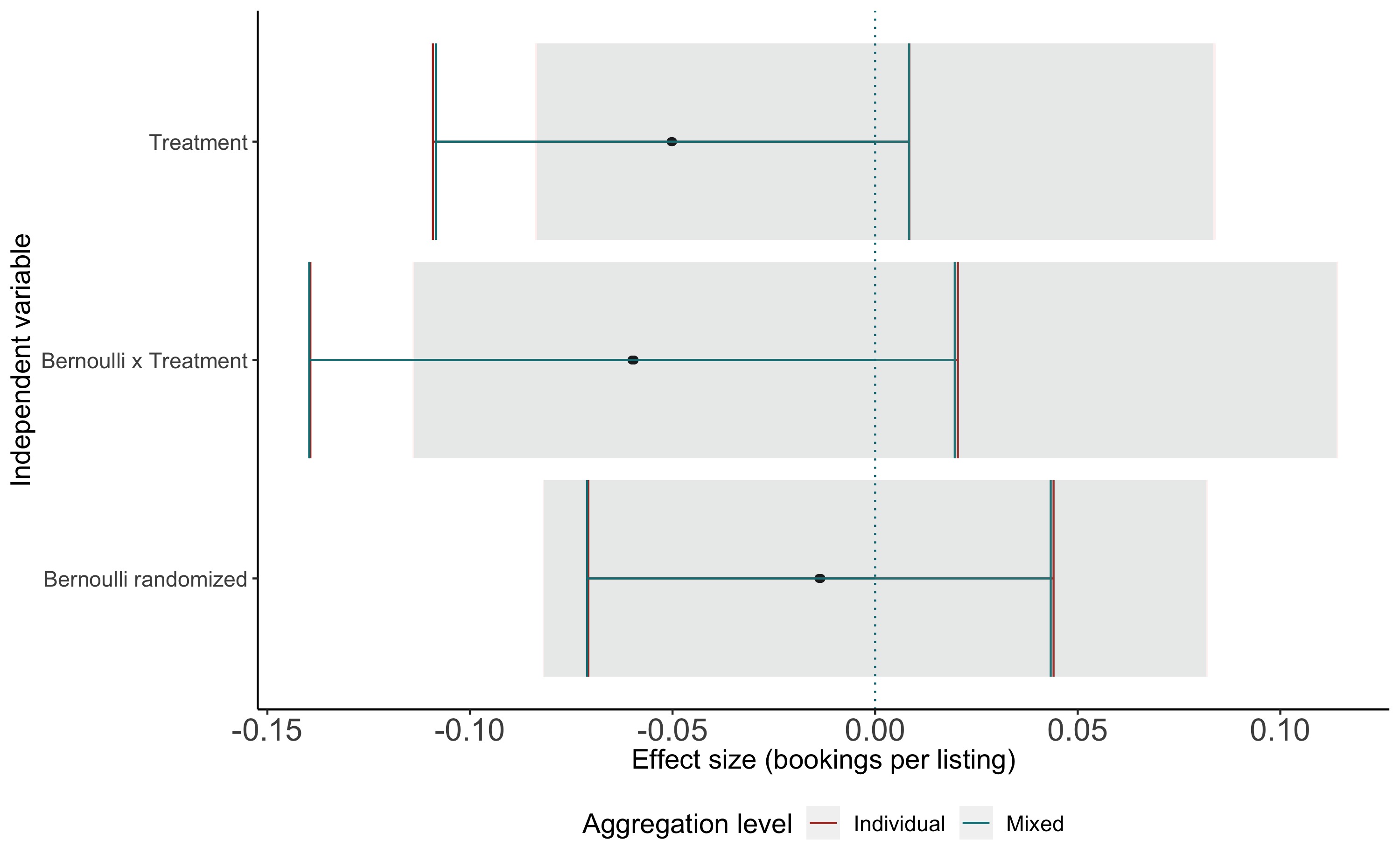}
      \caption{Comparison of algorithmic pricing experiment meta-analysis estimates obtained analyzing data at the individual level of analysis, and the mixed level of analysis. In the mixed analysis, Data from Bernoulli randomized listings is included at the listing level, whereas data from cluster randomized listings is aggregated at the cluster level. Error bars correspond to 95\% confidence intervals. Shaded areas represent effect sizes below the MDE threshold (80\% power, 95\% confidence).}
      \label{fig:meta_analysis_plot_bookings_only_comp_agg_pricing}
\end{figure}

\clearpage
\section{Additional Tables}

\begin{table}[ht]
\centering
\caption{The ratio of demand capture for 1,000 listing threshold clusters and 250 listing threshold clusters,
using different demand capture metrics and user subpopulations.} 
\label{tab:cluster_size_comparison}
\begin{tabular}{llrrrrr}
  \hline
Single views? & Type of viewers & avg. cluster share & avg. HHI & \% over 67\% & \% over 75\% & \% over 90\% \\ 
  \hline
No & All & 1.32 & 1.36 & 2.36 & 2.46 & 2.38 \\ 
  No & Bookers & 1.38 & 1.43 & 2.48 & 2.59 & 2.50 \\ 
  Yes & All & 1.16 & 1.19 & 1.37 & 1.33 & 1.26 \\ 
  Yes & Bookers & 1.23 & 1.27 & 1.54 & 1.49 & 1.37 \\ 
   \hline
\end{tabular}
\end{table}

\begin{table}[!htbp] \centering 
  \caption{Independent results of the fee meta-experiment (nights booked and gross guest spend)} 
  \label{tab:experiment_analysis_fees_no_bookings} 
\footnotesize 
\begin{tabular}{@{\extracolsep{5pt}}lcccc} 
\\[-1.8ex]\hline 
\hline \\[-1.8ex] 
 & \multicolumn{4}{c}{\textit{Dependent variable:}} \\ 
\cline{2-5} 
\\[-1.8ex] & \multicolumn{2}{c}{Nights booked} & \multicolumn{2}{c}{Gross guest spend} \\ 
 & Bernoulli randomized & Cluster randomized & Bernoulli randomized & Cluster randomized \\ 
\\[-1.8ex] & (1) & (2) & (3) & (4)\\ 
\hline \\[-1.8ex] 
 Treatment & $-$0.768$^{***}$ & $-$0.579$^{***}$ & $-$79.677$^{***}$ & $-$63.388$^{***}$ \\ 
  & (0.062) & (0.052) & (8.044) & (7.741) \\ 
  & & & & \\ 
 Pre-treatment bookings & 0.281$^{***}$ & 0.288$^{***}$ & 23.220$^{***}$ & 22.626$^{***}$ \\ 
  & (0.005) & (0.003) & (0.750) & (0.372) \\ 
  & & & & \\ 
 Pre-treatment nights booked & 0.038$^{***}$ & 0.037$^{***}$ & $-$4.289$^{***}$ & $-$3.698$^{***}$ \\ 
  & (0.002) & (0.001) & (0.433) & (0.129) \\ 
  & & & & \\ 
 Pre-treatment booking value & $-$0.000$^{***}$ & $-$0.000$^{***}$ & $-$0.060 & $-$0.148$^{***}$ \\ 
  & (0.000) & (0.000) & (0.085) & (0.021) \\ 
  & & & & \\ 
 Pre-treatment gross guest spend & 0.000$^{***}$ & 0.000$^{***}$ & 0.153$^{**}$ & 0.226$^{***}$ \\ 
  & (0.000) & (0.000) & (0.070) & (0.017) \\ 
  & & & & \\ 
\hline \\[-1.8ex] 
Stratum F.E. & Yes & Yes & Yes & Yes \\ 
Robust s.e. & Yes & Yes & Yes & Yes \\ 
Clustered s.e. & No & Yes & No & Yes \\ 
R$^{2}$ & 0.115 & 0.118 & 0.166 & 0.176 \\ 
Adjusted R$^{2}$ & 0.114 & 0.118 & 0.165 & 0.176 \\ 
\hline 
\hline \\[-1.8ex] 
\textit{Note:}  & \multicolumn{4}{r}{$^{*}$p$<$0.1; $^{**}$p$<$0.05; $^{***}$p$<$0.01} \\ 
\end{tabular} 
\end{table}

\begin{table}[!htbp] \centering 
  \caption{Results of the fees Meta-experiment (nights booked and gross guest spend)} 
  \label{tab:meta_experiment_analysis_fees_no_bookings} 
\begin{tabular}{@{\extracolsep{5pt}}lcc} 
\\[-1.8ex]\hline 
\hline \\[-1.8ex] 
 & \multicolumn{2}{c}{\textit{Dependent variable:}} \\ 
\cline{2-3} 
\\[-1.8ex] & Nights booked & Gross guest spend \\ 
\\[-1.8ex] & (1) & (2)\\ 
\hline \\[-1.8ex] 
 Treatment & $-$0.579$^{***}$ & $-$62.696$^{***}$ \\ 
  & (0.052) & (7.749) \\ 
  & & \\ 
 Bernoulli Randomized & 0.111 & 18.063$^{*}$ \\ 
  & (0.069) & (10.217) \\ 
  & & \\ 
 Bernoulli $\times$ Treatment & $-$0.191$^{**}$ & $-$16.704 \\ 
  & (0.082) & (11.085) \\ 
  & & \\ 
 Pre-treatment bookings & 0.287$^{***}$ & 22.787$^{***}$ \\ 
  & (0.002) & (0.342) \\ 
  & & \\ 
 Pre-treatment nights booked & 0.038$^{***}$ & $-$3.849$^{***}$ \\ 
  & (0.001) & (0.147) \\ 
  & & \\ 
 Pre-treatment booking value & $-$0.000$^{***}$ & $-$0.123$^{***}$ \\ 
  & (0.000) & (0.028) \\ 
  & & \\ 
 Pre-treatment gross guest spend & 0.000$^{***}$ & 0.206$^{***}$ \\ 
  & (0.000) & (0.023) \\ 
  & & \\ 
\hline \\[-1.8ex] 
Stratum F.E. & Yes & Yes \\ 
Robust s.e. & Yes & Yes \\ 
Clustered s.e. & Yes & Yes \\ 
R$^{2}$ & 0.117 & 0.173 \\ 
Adjusted R$^{2}$ & 0.117 & 0.173 \\ 
\hline 
\hline \\[-1.8ex] 
\textit{Note:}  & \multicolumn{2}{r}{$^{*}$p$<$0.1; $^{**}$p$<$0.05; $^{***}$p$<$0.01} \\ 
\end{tabular} 
\end{table}

\begin{table}[!htbp] \centering 
  \caption{Independent results of the algorithmic pricing meta-experiment (nights booked and gross guest spend)} 
  \label{tab:experiment_analysis_pricing_no_bookings} 
\footnotesize 
\begin{tabular}{@{\extracolsep{5pt}}lcccc} 
\\[-1.8ex]\hline 
\hline \\[-1.8ex] 
 & \multicolumn{4}{c}{\textit{Dependent variable:}} \\ 
\cline{2-5} 
\\[-1.8ex] & \multicolumn{2}{c}{Nights booked} & \multicolumn{2}{c}{Gross guest spend} \\ 
 & Bernoulli randomized & Cluster randomized & Bernoulli randomized & Cluster randomized \\ 
\\[-1.8ex] & (1) & (2) & (3) & (4)\\ 
\hline \\[-1.8ex] 
 Treatment & $-$0.288$^{**}$ & $-$0.176 & $-$37.377$^{**}$ & 2.268 \\ 
  & (0.139) & (0.118) & (17.052) & (16.466) \\ 
  & & & & \\ 
 Pre-treatment bookings & 1.342$^{***}$ & 1.370$^{***}$ & 87.218$^{***}$ & 85.714$^{***}$ \\ 
  & (0.013) & (0.008) & (1.842) & (1.095) \\ 
  & & & & \\ 
 Pre-treatment nights booked & 0.152$^{***}$ & 0.147$^{***}$ & $-$19.907$^{***}$ & $-$19.948$^{***}$ \\ 
  & (0.004) & (0.003) & (0.963) & (0.471) \\ 
  & & & & \\ 
 Pre-treatment booking value & $-$0.006$^{***}$ & $-$0.006$^{***}$ & $-$1.782$^{***}$ & $-$1.722$^{***}$ \\ 
  & (0.000) & (0.000) & (0.168) & (0.091) \\ 
  & & & & \\ 
 Pre-treatment gross guest spend & 0.005$^{***}$ & 0.005$^{***}$ & 2.083$^{***}$ & 2.038$^{***}$ \\ 
  & (0.000) & (0.000) & (0.141) & (0.078) \\ 
  & & & & \\ 
 Smart pricing pre-treatment & 3.376$^{***}$ & 3.437$^{***}$ & 362.779$^{***}$ & 348.078$^{***}$ \\ 
  & (0.164) & (0.096) & (23.840) & (13.857) \\ 
  & & & & \\ 
\hline \\[-1.8ex] 
Stratum F.E. & Yes & Yes & Yes & Yes \\ 
Robust s.e. & Yes & Yes & Yes & Yes \\ 
Clustered s.e. & No & Yes & No & Yes \\ 
R$^{2}$ & 0.282 & 0.283 & 0.381 & 0.373 \\ 
Adjusted R$^{2}$ & 0.280 & 0.282 & 0.379 & 0.373 \\ 
\hline 
\hline \\[-1.8ex] 
\textit{Note:}  & \multicolumn{4}{r}{$^{*}$p$<$0.1; $^{**}$p$<$0.05; $^{***}$p$<$0.01} \\ 
\end{tabular} 
\end{table}

\begin{table}[!htbp] \centering 
  \caption{Results of the algorithmic pricing meta-experiment (nights booked and gross guest spend)} 
  \label{tab:meta_experiment_analysis_pricing_no_bookings} 
\begin{tabular}{@{\extracolsep{5pt}}lcc} 
\\[-1.8ex]\hline 
\hline \\[-1.8ex] 
 & \multicolumn{2}{c}{\textit{Dependent variable:}} \\ 
\cline{2-3} 
\\[-1.8ex] & Nights booked & Booking value \\ 
\\[-1.8ex] & (1) & (2)\\ 
\hline \\[-1.8ex] 
 Treatment & $-$0.178 & 1.682 \\ 
  & (0.121) & (16.904) \\ 
  & & \\ 
 Bernoulli Randomized & $-$0.057 & 15.840 \\ 
  & (0.154) & (20.941) \\ 
  & & \\ 
 Bernoulli $\times$ Treatment & $-$0.104 & $-$37.238 \\ 
  & (0.184) & (23.988) \\ 
  & & \\ 
 Pre-treatment bookings & 1.366$^{***}$ & 86.295$^{***}$ \\ 
  & (0.007) & (0.941) \\ 
  & & \\ 
 Pre-treatment nights booked & 0.149$^{***}$ & $-$20.025$^{***}$ \\ 
  & (0.002) & (0.429) \\ 
  & & \\ 
 Pre-treatment booking value & $-$0.005$^{***}$ & $-$1.717$^{***}$ \\ 
  & (0.000) & (0.080) \\ 
  & & \\ 
 Pre-treatment gross guest spend & 0.005$^{***}$ & 2.033$^{***}$ \\ 
  & (0.000) & (0.068) \\ 
  & & \\ 
 Smart pricing pre-treatment & 3.382$^{***}$ & 344.350$^{***}$ \\ 
  & (0.084) & (12.096) \\ 
  & & \\ 
\hline \\[-1.8ex] 
Stratum F.E. & Yes & Yes \\ 
Robust s.e. & Yes & Yes \\ 
Clustered s.e. & Yes & Yes \\ 
R$^{2}$ & 0.281 & 0.374 \\ 
Adjusted R$^{2}$ & 0.280 & 0.373 \\ 
\hline 
\hline \\[-1.8ex] 
\textit{Note:}  & \multicolumn{2}{r}{$^{*}$p$<$0.1; $^{**}$p$<$0.05; $^{***}$p$<$0.01} \\ 
\end{tabular} 
\end{table}

\begin{table}[!htbp] \centering 
  \caption{Cluster randomized fees experiment (individual- and cluster-level analysis)} 
  \label{tab:cluster_experiment_analysis_fees_compare} 
\begin{tabular}{@{\extracolsep{5pt}}lcc} 
\\[-1.8ex]\hline 
\hline \\[-1.8ex] 
 & \multicolumn{2}{c}{\textit{Dependent variable:}} \\ 
\cline{2-3} 
 & Individual-level & Cluster-level \\ 
\\[-1.8ex] & (1) & (2)\\ 
\hline \\[-1.8ex] 
 Treatment & $-$0.142$^{***}$ & $-$0.137$^{***}$ \\ 
  & (0.011) & (0.011) \\ 
  & & \\ 
 Pre-treatment bookings & 0.174$^{***}$ & 0.206$^{***}$ \\ 
  & (0.001) & (0.006) \\ 
  & & \\ 
 Pre-treatment nights booked & $-$0.003$^{***}$ & 0.003$^{*}$ \\ 
  & (0.000) & (0.002) \\ 
  & & \\ 
 Pre-treatment booking value & 0.000$^{***}$ & $-$0.000 \\ 
  & (0.000) & (0.000) \\ 
  & & \\ 
 Pre-treatment gross guest spend & $-$0.000$^{***}$ & 0.000 \\ 
  & (0.000) & (0.000) \\ 
  & & \\ 
\hline \\[-1.8ex] 
Stratum F.E. & Yes & Yes \\ 
Robust s.e. & Yes & Yes \\ 
Clustered s.e. & Yes & No \\ 
R$^{2}$ & 0.405 & 0.973 \\ 
Adjusted R$^{2}$ & 0.405 & 0.968 \\ 
\hline 
\hline \\[-1.8ex] 
\textit{Note:}  & \multicolumn{2}{r}{$^{*}$p$<$0.1; $^{**}$p$<$0.05; $^{***}$p$<$0.01} \\ 
\end{tabular} 
\end{table}

\begin{table}[!htbp] \centering 
  \caption{Cluster randomized algorithmic pricing experiment (individual- and cluster-level analysis)} 
  \label{tab:cluster_experiment_analysis_pricing_compare} 
\begin{tabular}{@{\extracolsep{5pt}}lcc} 
\\[-1.8ex]\hline 
\hline \\[-1.8ex] 
 & \multicolumn{2}{c}{\textit{Dependent variable:}} \\ 
\cline{2-3} 
 & Individual-level & Cluster-level \\ 
\\[-1.8ex] & (1) & (2)\\ 
\hline \\[-1.8ex] 
 Treatment & $-$0.051$^{*}$ & $-$0.051$^{*}$ \\ 
  & (0.029) & (0.029) \\ 
  & & \\ 
 Pre-treatment bookings & 0.828$^{***}$ & 1.114$^{***}$ \\ 
  & (0.002) & (0.017) \\ 
  & & \\ 
 Pre-treatment nights booked & $-$0.017$^{***}$ & $-$0.006 \\ 
  & (0.000) & (0.005) \\ 
  & & \\ 
 Pre-treatment booking value & 0.000$^{***}$ & 0.000$^{**}$ \\ 
  & (0.000) & (0.000) \\ 
  & & \\ 
 Pre-treatment gross guest spend & $-$0.000$^{***}$ & $-$0.000$^{*}$ \\ 
  & (0.000) & (0.000) \\ 
  & & \\ 
 Smart pricing pre-treatment & 0.586$^{***}$ & $-$0.777$^{***}$ \\ 
  & (0.020) & (0.172) \\ 
  & & \\ 
\hline \\[-1.8ex] 
Stratum F.E. & Yes & Yes \\ 
Robust s.e. & Yes & Yes \\ 
Clustered s.e. & Yes & No \\ 
R$^{2}$ & 0.578 & 0.951 \\ 
Adjusted R$^{2}$ & 0.578 & 0.941 \\ 
\hline 
\hline \\[-1.8ex] 
\textit{Note:}  & \multicolumn{2}{r}{$^{*}$p$<$0.1; $^{**}$p$<$0.05; $^{***}$p$<$0.01} \\ 
\end{tabular} 
\end{table}

\begin{table}[!htbp] \centering 
  \caption{Results of the fees Meta-experiment (individual and mixed analysis)} 
  \label{tab:meta_experiment_analysis_fees_bookings_only_agg_comp} 
\begin{tabular}{@{\extracolsep{5pt}}lcc} 
\\[-1.8ex]\hline 
\hline \\[-1.8ex] 
 & \multicolumn{2}{c}{\textit{Dependent variable:}} \\ 
\cline{2-3} 
\\[-1.8ex] & \multicolumn{2}{c}{Bookings} \\ 
\\[-1.8ex] & (1) & (2)\\ 
\hline \\[-1.8ex] 
 Treatment & $-$0.139$^{***}$ & $-$0.139$^{***}$ \\ 
  & (0.011) & (0.011) \\ 
  & & \\ 
 Bernoulli Randomized & 0.022 & 0.021 \\ 
  & (0.014) & (0.014) \\ 
  & & \\ 
 Bernoulli $\times$ Treatment & $-$0.067$^{***}$ & $-$0.068$^{***}$ \\ 
  & (0.016) & (0.016) \\ 
  & & \\ 
 Pre-treatment bookings & 0.174$^{***}$ & 0.175$^{***}$ \\ 
  & (0.001) & (0.001) \\ 
  & & \\ 
 Pre-treatment nights booked & $-$0.003$^{***}$ & $-$0.003$^{***}$ \\ 
  & (0.000) & (0.000) \\ 
  & & \\ 
 Pre-treatment booking value & 0.000$^{***}$ & 0.000 \\ 
  & (0.000) & (0.000) \\ 
  & & \\ 
 Pre-treatment gross guest spend & $-$0.000$^{***}$ & $-$0.000 \\ 
  & (0.000) & (0.000) \\ 
  & & \\ 
\hline \\[-1.8ex] 
Stratum F.E. & Yes & Yes \\ 
Robust s.e. & Yes & Yes \\ 
Clustered s.e. & Yes & No \\ 
R$^{2}$ & 0.405 & 0.515 \\ 
Adjusted R$^{2}$ & 0.405 & 0.515 \\ 
\hline 
\hline \\[-1.8ex] 
\textit{Note:}  & \multicolumn{2}{r}{$^{*}$p$<$0.1; $^{**}$p$<$0.05; $^{***}$p$<$0.01} \\ 
\end{tabular} 
\end{table}

\begin{table}[!htbp] \centering 
  \caption{Results of the algorithmic pricing meta-experiment (individual and mixed analysis)} 
  \label{tab:meta_experiment_analysis_pricing_bookings_only_agg_comp} 
\begin{tabular}{@{\extracolsep{5pt}}lcc} 
\\[-1.8ex]\hline 
\hline \\[-1.8ex] 
 & \multicolumn{2}{c}{\textit{Dependent variable:}} \\ 
\cline{2-3} 
\\[-1.8ex] & \multicolumn{2}{c}{Bookings} \\ 
\\[-1.8ex] & (1) & (2)\\ 
\hline \\[-1.8ex] 
 Treatment & $-$0.050$^{*}$ & $-$0.050$^{*}$ \\ 
  & (0.030) & (0.030) \\ 
  & & \\ 
 Bernoulli Randomized & $-$0.013 & $-$0.014 \\ 
  & (0.037) & (0.037) \\ 
  & & \\ 
 Bernoulli $\times$ Treatment & $-$0.059 & $-$0.060 \\ 
  & (0.041) & (0.041) \\ 
  & & \\ 
 Pre-treatment bookings & 0.827$^{***}$ & 0.838$^{***}$ \\ 
  & (0.002) & (0.004) \\ 
  & & \\ 
 Pre-treatment nights booked & $-$0.017$^{***}$ & $-$0.018$^{***}$ \\ 
  & (0.000) & (0.001) \\ 
  & & \\ 
 Pre-treatment booking value & 0.000$^{***}$ & 0.000 \\ 
  & (0.000) & (0.000) \\ 
  & & \\ 
 Pre-treatment gross guest spend & $-$0.000$^{***}$ & $-$0.000$^{*}$ \\ 
  & (0.000) & (0.000) \\ 
  & & \\ 
 Smart pricing pre-treatment & 0.577$^{***}$ & 0.358$^{***}$ \\ 
  & (0.017) & (0.037) \\ 
  & & \\ 
\hline \\[-1.8ex] 
Stratum F.E. & Yes & Yes \\ 
Robust s.e. & Yes & Yes \\ 
Clustered s.e. & Yes & No \\ 
R$^{2}$ & 0.577 & 0.692 \\ 
Adjusted R$^{2}$ & 0.577 & 0.691 \\ 
\hline 
\hline \\[-1.8ex] 
\textit{Note:}  & \multicolumn{2}{r}{$^{*}$p$<$0.1; $^{**}$p$<$0.05; $^{***}$p$<$0.01} \\ 
\end{tabular} 
\end{table}

\end{APPENDICES}

\clearpage

\ACKNOWLEDGMENT{The authors are grateful to Lanbo Zhang, Minyong Lee, and Sharan Srinivasan for their assistance with the design and analysis of the experiments in this paper. We also thank numerous other Airbnb employees who have assisted with this project. We also appreciate the helpful feedback we have received from Dean Eckles, Andrey Fradkin, Alex Moehring, Hong Yi Tu Ye, attendees of the 2019 Winter Conference on Business Analytics and the HBS Digital Doctoral Workshop. This experiment was classified as exempt by the MIT Committee on the Use of Humans as Experimental Subjects under Protocol \#1807452488.}





\end{document}